\documentclass[11pt,a4paper]{article}
\pdfoutput=1
\usepackage{jheppub}
\usepackage{amsmath}
\usepackage{dsfont}
\usepackage{ulem}
\usepackage{natbib}
\usepackage{xcolor}
\usepackage[hang,flushmargin,bottom,splitrule]{footmisc}
\usepackage{tikz-cd}
\usepackage{array}
\usepackage{cancel}

\makeatletter\renewcommand{\@biblabel}[1]{#1.}\makeatother

\def\be{\begin{equation}}
\def\ee{\end{equation}}
\def\bes{\begin{subequations}}
\def\ees{\end{subequations}}

\def\vep{\varepsilon}
\def\i{{\rm i}}
\def\e{{\rm e}}
\def\d{{\rm d}}
\def\ul{\underline}

\def\nn{{\nonumber}}
\def\q{{\mathfrak{q}}}
\def\t{{\mathfrak{t}}}
\def\p{{\mathfrak{p}}}
\def\L{{\text{L}}}
\def\R{{\text{R}}}
\newcommand{\mb}[1]{\mathbf{#1}}

\def\ket#1{{ |#1\rangle}}
\def\bra#1{{ \langle#1|}}
\def\braket #1#2{{ \langle #1|#2 \rangle}}

\preprint{{\small{\textsf{DESY 21-060}}}}

\title{
Intersecting Defects and Supergroup Gauge Theory
}

\author[a]{Taro Kimura}

\author[b]{and Fabrizio Nieri}

\affiliation[a]{Institut de Math{\'e}matiques de Bourgogne\\ Universit{\'e} Bourgogne Franche-Comt{\'e}, 21078 Dijon, France.}

\affiliation[b]{DESY Theory Group\\
Notkestraße 85, 22607 Hamburg, Germany.}

\emailAdd{taro.kimura@u-bourgogne.fr}

\emailAdd{fb.nieri@gmail.com}

\abstract{We consider 5d supersymmetric gauge theories with unitary groups in the $\Omega$-background and study codim-2/4 BPS defects supported on orthogonal planes intersecting at the origin along a circle. The intersecting defects arise upon implementing the most generic Higgsing (geometric transition) to the parent higher dimensional theory, and they are described by pairs of 3d supersymmetric gauge theories with unitary groups interacting through 1d matter at the intersection. We explore the relations between instanton and generalized vortex calculus, pointing out a duality between intersecting defects subject to the $\Omega$-background and a deformation of supergroup gauge theories, the exact supergroup point being achieved in the self-dual or unrefined limit. Embedding our setup into refined topological strings and in the simplest case when the parent 5d theory is Abelian, we are able to identify the supergroup theory dual to the intersecting defects as the supergroup version of refined Chern-Simons theory via open/closed duality. We also discuss the BPS/CFT side of the correspondence, finding an interesting large rank duality with super-instanton counting.}


\begin{document}

\maketitle
\newpage


\section{Introduction}

The models of physics are built on symmetry principles, which can in turn be exploited to organize their phenomenology. Continuous symmetries are described by Lie groups and algebras, however, it was long recognized that the generalization to supergroups and superalgebras, comprising fermionic or odd generators in addition to the bosonic or even ones, are also necessary in order to not miss out on important class of theories. Fundamental examples are string theory and supersymmetric gauge theories. Often, superalgebras appear as peculiar global (as opposed to local) symmetries dictating the pairing between bosonic and fermionic degrees of freedom. 

\paragraph{Why gauge supergroups?} Since supergroups provide a natural unifying opportunity treating particles of opposite statistics quite democratically, while gauge invariance represents the main model building principle for our comprehension of nature, it is natural to wonder what is the position occupied by QFT (supersymmetric or not) based on gauge supergroups. Non-unitarity is obvious due to the violation of the spin-statistics theorem. Also, the lack of a definite bilinear form on a Lie superalgebra poses some question about their very quantum definition and consistency, usually requiring some non-perturbative approach. Despite such ``unattractive'' features, supergroup gauge theories and related supergroup matrix models inevitably show up in many places of theoretical physics, string/M-theory and brane dynamics in particular. This is perhaps not surprising given the intrinsic non-perturbative nature of such models. Just to mention few remarkable examples, the supergroup structure of the ABJ(M) partition function (the theory holographic dual to M-theory on $\text{AdS}_4$) \cite{Aharony:2008ug,Aharony:2008gk,Kapustin:2009kz} was instrumental for constructing $1/2$ BPS Wilson loops \cite{Drukker:2009hy}. The supermatrix model description was also crucial for exact investigations of non-perturbative aspects, in connection with holography and topological strings via Chern-Simons theory \cite{Drukker:2010nc,Drukker:2011zy,Marino:2011eh,Hatsuda:2013oxa} and open/closed large rank duality \cite{Gopakumar:1998ki}. Moreover, exotic and very interesting phenomena such as non-unitary holography and dynamical changes of space-time signature were also ascribed to supergroup gauge theories \cite{Vafa:2014iua,Dijkgraaf:2016lym}, whose realization within string theory can be achieved through the introduction of negative branes \cite{Vafa:2001qf,Okuda:2006fb} (essentially differing from anti-branes by their negative tension). More recently, it was shown that equivariant instanton calculus can be extended to supersymmetric gauge theories in the supergroup setting \cite{Kimura:2019msw}, revealing also far reaching connections to ``ordinary'' theories (in the sense of usual bosonic gauge groups) via integrability \cite{Chen:2019vvt,Chen:2020rxu,Nekrasov:2018gne}. Therefore, the main message past and recent investigations are telling us is that supergroup gauge theories, if properly understood quantum mechanically, are capable of shedding some light on poorly understood phenomena, or can effectively capture some corner of more traditional setups. In this regard, the modern developments of exact techniques in supersymmetric QFT \cite{Pestun:2016zxk} and the discoveries of non-perturbative duality webs \cite{Alday:2009aq,Awata:2009ur,Dimofte:2011ju}, represent an opportunity and a motivation for exploring supergroup gauge theories and the closely related supermatrix models (including deformations thereof) to learn about seemingly unrelated theories.

\paragraph{Intersecting defects.} Indeed, in this paper we consider yet another extension of conventional (supersymmetric) gauge theories, namely those based on ordinary gauge groups but supported on intersecting subspaces embedded in some ambient smooth manifold. The general and systematic definition of such theories involves the gauge origami construction introduced in \cite{Nekrasov:2016ydq,Nekrasov:2016gud}. Despite the complicated nature of such theories, the amount of supersymmetry they preserve makes them amenable of exact computations via localization. More specifically, our focus is actually on a much simpler setup than the full higher dimensional gauge origami background: the intersecting gauge theory arises as a codim-2/4 BPS defect upon Higgsing \cite{Nekrasov:2017rqy} an ordinary parent 5d $\mathcal{N}=1$ gauge theory in the $\Omega$-background $\mathbb{C}^2_{\q,\t^{-1}}\times \mathbb{S}^1$. For concreteness, we consider $\text{U}(N)$ SQCD only, the reason being that the defect ends up coupled to a free bulk, considerably simplifying explicit computations (quiver generalizations are certainly interesting). The support of the  defect is given by the two orthogonal cigars $\mathbb{C}_\q\times \mathbb{S}^1$ and $\mathbb{C}_{\t^{-1}}\times \mathbb{S}^1$, intersecting at the origin along a common $\mathbb{S}^1$. This description allows us to compute the defect partition function by implementing the most generic Higgsing condition directly on the instanton partition function of the parent theory. 

\paragraph{Summary of the main results.} Our analysis suggests a natural connection between (deformations of) supergroup gauge theories supported on ordinary manifolds and supersymmetric theories with ordinary gauge groups supported on intersecting ($\Omega$-deformed) spaces. This observation is the main point of the paper and an additional motivation for studying and developing supergroup gauge theories. We argue that the defect on the intersecting cigars provide a natural $(\q,\t)$-deformation of a dual supergroup gauge theory, supporting this view by studying the resulting partition function in the holomorphic block representation \cite{Beem:2012mb} (i.e. Neumann conditions for the vectors \cite{Yoshida:2014ssa}). This is shown to give a $(\q,\t)$-deformation of a supergroup matrix model, with the honest supergroup structure appearing in the so-called unrefined limit $\t=\q$. In principle, diverse deformations of supermatrix models may exist: our results fit with the deformation considered in \cite{Atai_2019}. In a particular example, namely a partially matter decoupling limit applied to SQED, we can identify the dual supergroup gauge theory as the supergroup version of refined Chern-Simons theory on $\mathbb{S}^3$ \cite{Aganagic:2011sg,Aganagic:2012hs}. Quite interestingly, our results also suggest that the refinement can be naturally understood as a specific deformation away from the supergroup point, primarily detected by two unrelated couplings encoded in $\t\neq \q$. This perspective completes rather ad-hoc setups explicitly breaking the $\q\leftrightarrow \t^{-1}$ symmetry of the $\Omega$-background. We can observe that the appearance of a supergroup structure (exact in the unrefined limit only) is quite obscure from the intersecting defect perspective: the parent theory is an ordinary gauge theory, while after Higgsing the elementary degrees of freedom are even defined on different space-time components. Nevertheless,  an explanation can be provided by mapping our configuration to a Mikhaylov-Witten construction \cite{Mikhaylov:2014aoa} via a chain of dualities in string/M-theory. Similar arguments have recently appeared in the context of categorification of supergroup Chern-Simons observables \cite{Ferrari:2020avq}.

The rest of this paper is organized as follows. In section \ref{sec:2}, we consider the most generic Higgsing procedure applied to 5d SQCD, which gives rise to the intersecting defect of our interest. We present the derivation of the 3d-1d coupled holomorphic blocks in different chambers of the $\Omega$-background parameters, and discuss the equivalence with a specific $(\q,\t)$-deformation of the supergroup matrix model, also arising in the context of relativistic many-body integrable systems. In section \ref{sec:3}, we provide an explanation of the emerging supergroup structure of the defect partition function by considering a chain of string dualities, including the embedding into (refined) topological strings and open/closed large rank duality. This perspective yields, to some extent, a generalization of the 3d-3d correspondence to 3d(intersecting)-3d(supergroup) theories. In section \ref{sec:4}, we discuss the BPS/CFT side of the story, namely the doubly quantized Seiberg-Witten geometry and the associated $(\q,\t)$-deformed W algebras. In the unrefined limit, our analysis reveals (unexpected) connections to supergroup instanton calculus. Finally, in section \ref{sec:5}, we comment on open issues and directions for future investigations. The paper is supplemented by an appendix where conventions and more technical computations as reported.

\section{Intersecting defects from Higgsing}\label{sec:2}

In this section, we consider the 5d $\mathcal{N}=1$ SYM theory with gauge group $\text{U}(N)$ coupled to $N$ fundamental and $N$ anti-fundamental hypers (SQCD). We then impose a degeneration condition between the Coulomb and fundamental mass parameters to reach the root of the Higgs branch, where the defect description opens up. Generically, the Higgsed theory reduces to a 5d-3d-1d coupled gauge system, but in our case the 5d gauge group is completely frozen in the procedure. We are thus left with a 3d-1d defect coupled to a free bulk, and the non-trivial vortex-like part of the partition function coincides with the specialization of the instanton partition function. Compact and non-compact space versions of our setup were already studied in several papers \cite{Gomis:2016ljm,Pan:2016fbl,Nieri:2017ntx,Nieri:2018pev}. The deformation parameters $\q$ and $\t^{-1}$ are related to the usual 4d conventions as follows
\be
\q\equiv \e^{- R\varepsilon_1}~,\quad \t^{-1}\equiv \e^{-R\varepsilon_2}~,\quad \p\equiv\e^{- R\varepsilon_+}~,
\ee
where $R$ measures the size of the fifth compact dimension and $\varepsilon_+\equiv \varepsilon_1+\varepsilon_2$.  From the bulk perspective, it is quite natural to consider a chamber of the parameter space where $\q$ and $\t^{-1}$ appear on equal footing (e.g. $|\q|<1,|\t^{-1}|<1$) due to the manifest exchange symmetry ($\varepsilon_1\leftrightarrow\varepsilon_2$). However, for our purposes it is also convenient to consider a description adapted to the case where there is some democracy between $\q$ and $\t$ (e.g. $|\q|<1,|\t|<1$) as we are eventually interested in studying the self-dual or unrefined $\t=\q$ limit  ($\varepsilon_+=0$). In the first chamber, this limit happens at the boundary, namely on the unit circle, forcing  both parameters to be pure phases (as familiar from the Chern-Simons description); in the second chamber, the limit makes sense within the unit disk (hence providing an analytic continuation). We study both regimes and provide a more detailed derivation adapted to the case $|\q|<1,|\t|<1$, which was not considered in the literature before: as we will see, there are some subtleties to be understood. 

\subsection{Matrix model of the intersecting defect}

\paragraph{Hook factorization.} Let us start by recalling some basic results and definitions from K-theoretic instanton counting in 5d $\mathcal{N}=1$ quiver gauge theories with unitary gauge groups. After equivariant localization at the fixed points of the instanton moduli space, the instanton partition function can be written in a combinatorial form  \cite{Nekrasov:2002qd,Nekrasov:2003rj} where the Nekrasov function is the main building block (we use the conventions of \cite{Awata:2008ed})
\be
N_{\mu\nu}(x;\q,\t)\equiv \prod_{(i,j)\in\mu}(1-x \q^{\mu_i-j}\t^{\nu^\vee_j-i+1})\prod_{(i,j)\in\nu}(1-x \q^{-\nu_i+j-1}\t^{-\mu^\vee_j+i})~.
\ee
Here $\q,\t^{-1}$ are independent $\mathbb{C}^\times$ parameters identified with the rotational equivariant parameters of the $\Omega$-background $\mathbb{C}^2_{\q,\t^{-1}}\times \mathbb{S}^1$ as before, while $\mu$, $\nu$ represent integer partitions or Young diagrams (i.e. $\mu_{i}\geq \mu_{i+1}\geq 0$, $\nu_{i}\geq \nu_{i+1}\geq 0$) parametrized by the coordinates $(i,j)$ of boxes running over the rows and columns respectively, with ${}^\vee$ the transpose operation. The following identities relating the $\q\leftrightarrow \t^{-1}$ or $\q\leftrightarrow \t$ exchanges will be important
\begin{multline}\label{app:neksymm}
N_{\mu^\vee \nu^\vee}(x;\t^{-1},\q^{-1})=N_{\mu \nu}(x;\q,\t)=
N_{\nu^\vee \mu^\vee}(\p^{-1}x;\t,\q)=\\
=N_{\mu^\vee \nu^\vee}(x^{-1};\t,\q)\ (\p^{-1/2}x)^{|\mu|+|\nu|}f_{\nu^\vee}(\t,\q)/f_{\mu^\vee}(\t,\q)~,
\end{multline}
where $\p\equiv \q \t^{-1}$, while $|\mu|\equiv \sum_i\mu_i$ denotes the number of boxes and 
\be
f_\mu(\q,\t)\equiv \prod_{(i,j)\in\mu}-\q^{\mu_i-j+1/2}\t^{-\mu^\vee_j+i-1/2}\equiv(-1)^{|\mu|}\q^{\|\mu\|^2/2}\t^{-\|\mu^\vee\|^2/2}~,
\ee
is the framing factor, with $\|\mu\|^2\equiv \sum_i \mu_i^2$.

\textit{Remark.} In order to make contact with the 4d or cohomological notations, we must parametrize $\q\equiv \exp(-R\vep_1)$, $\t\equiv \exp(R\vep_2)$, $x\equiv  \exp(-R X)$, and let the 5d radius $R$ go to zero (while keeping fixed the instanton counting parameter). The unrefined limit $\t=\q$ corresponds to the limit $\p=1$, namely the self-dual point $\vep_+\equiv \vep_1+\vep_2=0$.

\begin{figure}[!ht]
\begin{center}
\includegraphics[width=0.9\textwidth]{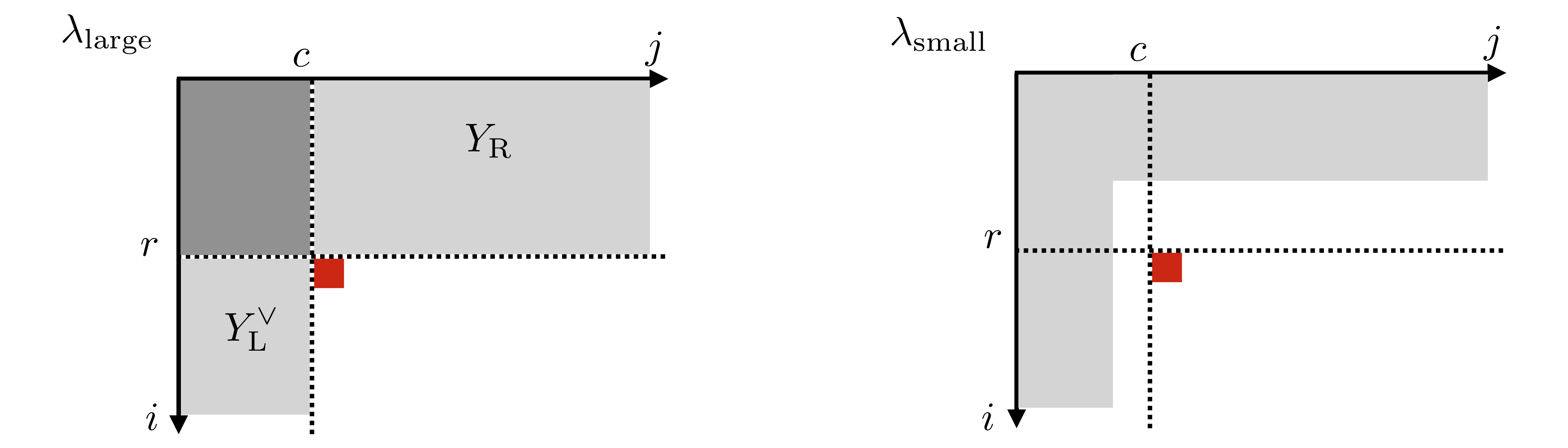}
\end{center}
\caption{Distinction between large (left side) and small (right side) hook diagrams of type $(r|c)$ avoiding box at $(i,j)=(r+1,c+1)$ (red) and their decomposition into Right (R) and Left (L) sub-diagrams (large only). Large diagrams have the rectangular region $r\times c$ completely filled (dark gray), and Right/Left diagrams (light gray) can be built on the plateau. Small diagrams are those entering the rectangular region.}
\label{fig:LargeSmall-RL}
\end{figure}

The Nekrasov function has several zeros at  specific values of the variable $x$. In particular, when one of the diagrams is trivial, for $r,c\in\mathbb{Z}_{\geq 0}$ we have

\be\label{eq:NekZeros}
N_{\mu\emptyset}(\q^{-c}\t^{r};\q,\t)=0 \quad \text{if}\quad (r+1,c+1)\in \mu~,
\ee
and similarly for $N_{\emptyset\mu}(\p\q^{c}\t^{-r};\q,\t)$ by using the reflection properties listed above. This is most easily seen by using the expression
\be
N_{\mu\emptyset}(x;\q,\t)= \prod_{(i,j)\in\mu}(1-x\, \q^{\mu_i-j}\t^{-i+1})=\prod_{(i,j)\in\mu}(1-x\, \q^{j-1}\t^{-(i-1)})~.
\ee
This truncation is also known as the hook or pit condition \cite{Bershtein_2018} as only the diagrams which fit into the hook-shaped region avoiding the box $(r+1,c+1)$ give a non-vanishing result. Following \cite{Pan:2016fbl}, we refer to a diagram $\lambda$ of hook type $(r|c)$ as large if the upper-left $r\times c$ rectangle is completely filled (i.e. $\mu_{r}\geq c\geq \mu_{r+1}$), otherwise small (Fig.\ref{fig:LargeSmall-RL}). In the following, we review the factorization properties of the Nekrasov function when evaluated on large hook diagrams. 

Having in mind applications to $\text{U}(N)$ gauge theory, let us start by introducing a set of $N$ Young diagrams $\{\lambda_A~, A\in[1,N]\}$, and extract for one the maximal upper-left rectangle or plateau (i.e. we consider any given diagram as a large hook). We denote the number of rows and columns of the rectangle with $r_A, c_A$ respectively, so that 
\begin{align}
  \lambda_{Ai} \ge c_A, \quad i = 1, \ldots, r_A~,\qquad \lambda_{Ai} \le c_A, \quad i \geq r_A + 1 ~ .
\end{align}
We also set
\be
r\equiv \sum_A r_A~,\quad c\equiv \sum_A c_A~.
\ee
It is convenient to decompose the Young diagram $\lambda_A$ into $Y^\text{L}_A$ and $Y^\text{R}_A$ 
\begin{align}
  \lambda_{Ai} \equiv Y^\text{R}_{Ai}+c_A~,\quad i = 1, ..., r_A ~,\qquad \lambda_{A,r_A+i} \equiv Y^{\text{L}\vee}_{Ai}~,\quad i \geq 1 ~.
\end{align}
Note that $Y_A^{\text{R,L}}$ are sub-diagrams with at most $r_A,c_A$ rows respectively. We can now factorize $N_{\lambda_A \lambda_B}$ into Right and Left parts (times a remainder). The details of the derivation are given in appendix \ref{app:computations}, here we simply state the main result. For a set
\be
\ul x\equiv \{x_A~,A\in[1,N]\}
\ee
of $\mathbb{C}^\times$ Coulomb branch parameters, we have
\begin{multline}\label{eq:NekFac}
\prod_{A,B=1}^N \frac{1}{N_{\lambda_A \lambda_B}(x_A/x_B;\q,\t)}=\prod_{A,B=1}^N\frac{1}{N_{c_A^{r_A}c_B^{r_B}}(x_A/x_B;\q,\t)}\times\\
\times\p^{N| Y^\L|}\, \frac{\Upsilon_\text{v}(\ul z_{Y^\R};\q,\t)}{\Upsilon_\text{v}(\ul z_{\emptyset^\R};\q,\t)}\frac{\Upsilon_\text{v}(\ul z_{Y^\L};\t,\q)}{\Upsilon_\text{v}(\ul z_{\emptyset^L};\t,\q)}\times\\
\times\frac{\Upsilon_\text{x}(\eta_\R\, \ul x \q^{\ul c}\t^{-\ul r},\ul z_{Y^\R};\q,\t)}{\Upsilon_\text{x}(\eta_\R\, \ul x \q^{\ul c}\t^{-\ul r},\ul z_{\emptyset^\R};\q,\t)}\frac{\Upsilon_\text{x}(1/\eta_\L\, \ul x \q^{\ul c}\t^{-\ul r},1/\ul z_{Y^\L};\t,\q)}{\Upsilon_\text{x}(1/\eta_\L\, \ul x \q^{\ul c}\t^{-\ul r},1/\ul z_{\emptyset^\L};\t,\q)} \frac{\Upsilon^\text{1d}_\text{int}(\ul z_{Y^\L},\ul z_{Y^\R})}{\Upsilon^\text{1d}_\text{int}(\ul z_{\emptyset^\L},\ul z_{\emptyset^\R})}~,
\end{multline}
where we defined the dynamical variables 
\bes
\begin{align}
\ul z_{Y^\R}\equiv & \ \{\eta_\R x_A \q^{c_A}t^{-i}\q^{Y^\R_{Ai}}~, A\in [1,N]~, i \in [1,r_A]\}~,\\
\ul z_{Y^\L}\equiv & \ \{\eta_\L x_A \t^{-r_A}\q^{i}\t^{-Y^\L_{Ai}}~,A\in [1,N]~, i \in [1,c_A]\}~,
\end{align}
\ees
and for a set of variables $\ul u$, $\ul w$ we defined the functions
\bes
\begin{align}
\Upsilon_\text{v}(\ul u;\q,\t)\equiv & \ \prod_{i\neq j}\frac{(u_i/u_j;\q)_\infty}{(\t\, u_i/u_j;\q)_\infty}\equiv \Delta_\t(\ul u;\q)~,\label{eq:MacKer}\\
\Upsilon_\text{x}(\ul u,\ul w;\q,\t)\equiv & \ \prod_{i,j}\frac{(\t w_i/u_j;\q)_\infty}{(u_j/w_i;\q)_\infty}~,\\
\Upsilon^\text{1d}_\text{int}(\ul u ,\ul w)\equiv & \ \prod_{i,j}\frac{1}{(1-\q^{-1} \eta_\R \eta_\L^{-1} u_j/w_i)(1-\t\, \eta_\L \eta_\R^{-1} w_i/u_j)}~.
\end{align}
\ees
For later purposes, we have also introduced the Macdonald kernel in (\ref{eq:MacKer}). Note that in (\ref{eq:NekFac}) the $\mathbb{C}^\times$ parameters $\eta_{\L,\R}$ are just dummy constants. However, it may be convenient to fix this freedom and set $\eta_\R/\eta_\L\equiv \sqrt{\q\t}$.

\textit{Remark.} Nekrasov's summands are rational functions of $\q,\t$, thus they make sense for both the chambers $|\q|<1,|\t^{-1}|<1$ and $|\q|<1,|\t|<1$. The function to which the instanton sum will converge depends on this choice. While the gauge theory prefers the first chamber due to the intrinsic $\q\leftrightarrow \t^{-1}$ symmetry, we are here adapting to the second choice (mostly because it was not explicitly discussed in the literature in our context). We will comment on the differences w.r.t. the first chamber later on. 

The main feature of the formula above is that the interaction between the Left and Right sub-diagrams $Y^{\L,\R}$ is captured by the simple factor $\Upsilon_\text{int}$. We can use the very same result to also factorize the contribution due to (anti-)fundamental matter. For the case $N_\text{f}=N_\text{a.f}=N$, denoting by
\be
\ul\mu \equiv \{\mu_A~, A\in [1,N]\}~,\quad \ul{\bar\mu}\equiv \{\bar\mu_A~, A\in [1,N]\}
\ee
a set of $\mathbb{C}^\times$ flavor fugacities,\footnote{The bar is notation and does not denote complex conjugation.} we have
\begin{multline}
\prod_{A,B=1}^N N_{\emptyset \lambda_A}(\bar \mu_B/x_A;\q,\t)N_{\lambda_A \emptyset}(x_A/\mu_B;\q,\t)=\!\!\!\prod_{A,B=1}^N N_{\emptyset c_A^{r_A}}(\bar \mu_B/x_A;\q,\t)N_{c_A^{r_A} \emptyset}(x_A/\mu_B;\q,\t)\times\\
\times \frac{\Upsilon_\text{f}(\eta_\R\ul{\bar\mu},\eta_\R \ul\mu, \ul z_{Y^\R};\q,\t)}{\Upsilon_\text{f}(\eta_\R\ul{\bar\mu},\eta_\R \ul\mu, \ul z_{\emptyset^\R};\q,\t)}\frac{\Upsilon_\text{f}(1/\eta_\L\ul{\bar\mu},1/\eta_\L \ul\mu, 1/\ul z_{Y^\L};\t,\q)}{\Upsilon_\text{f}(1/\eta_\L\ul{\bar\mu},1/\eta_\L \ul\mu, 1/\ul z_{\emptyset^\L};\t,\q)}\times \p^{-N |Y^\L|}\prod_A (\bar \mu_A/\mu_A)^{|Y^\L |}~,
\end{multline}
where we defined the function
\be
\Upsilon_\text{f}(\ul{\bar u},\ul u,\ul w;\q,\t)\equiv \prod_{i,j}\frac{(\bar u_j/w_i;\q)_\infty}{(\t w_i/u_j;\q)_\infty}~.
\ee

\begin{figure}[!ht]
\begin{center}
\includegraphics[width=1\textwidth]{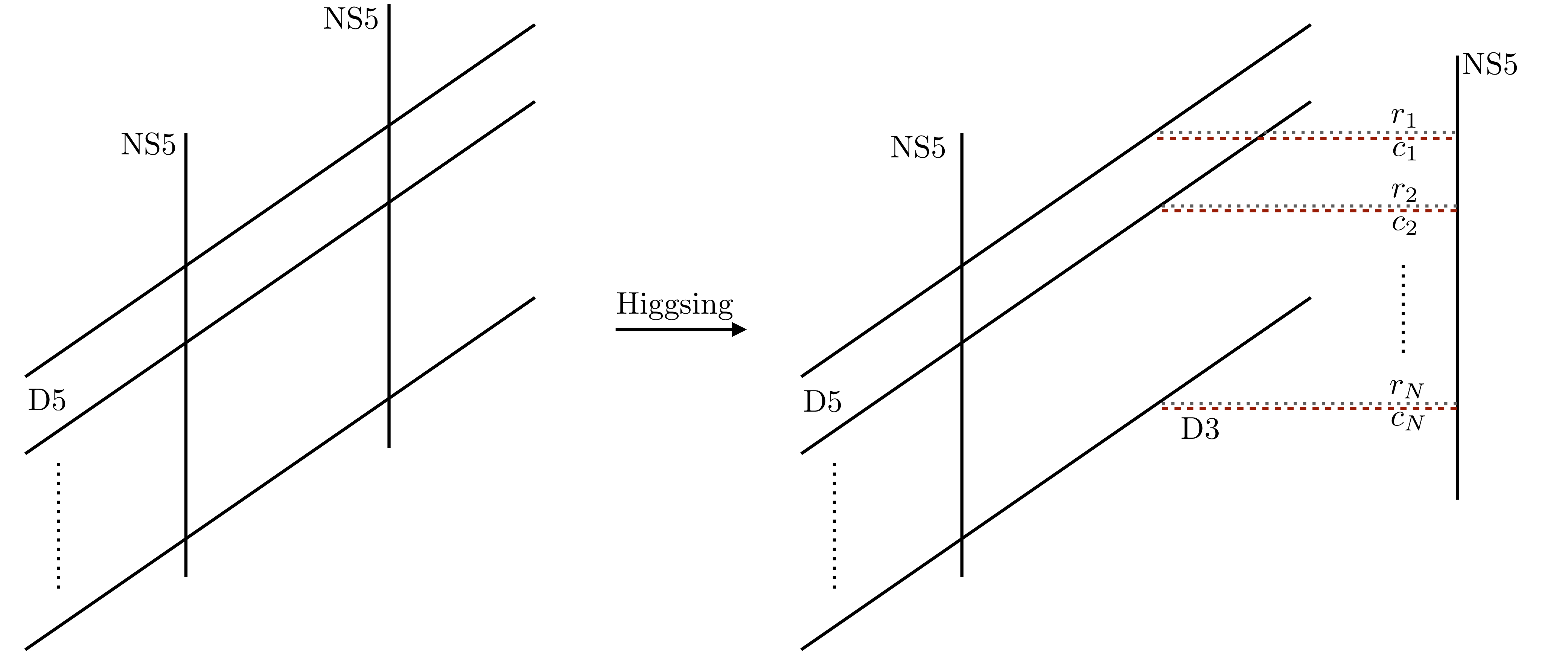}
\end{center}
\caption{The IIB $(p,q)$-web setup describing the 5d $\text{U}(N)$ SQCD (left side), and the Higgsing process with $r=\sum_{A=1}^N r_A$ and $c=\sum_{A=1}^N c_A$ stretched D3s extending along $\mathbb{C}_\q\times\mathbb{S}^1$ (dotted gray) and $\mathbb{C}_{\t^{-1}}\times\mathbb{S}^1$ (dashed red) respectively, supporting the intersecting defect theory (right side).}
\label{fig:pqDefect}
\end{figure}

\paragraph{Higgsing SQCD.} Given these preliminary results, we can now consider the K-theoretic SQCD instanton partition function, which using the previous notation reads as
\begin{align}
Z[\text{SQCD}]\equiv & \ \sum_{\{\lambda_A\}}\Lambda^{\sum_A|\lambda_A|}\, Z_{\{\lambda_A\}}[\text{SQCD}]~,\nn\\
Z_{\{\lambda_A\}}[\text{SQCD}]\equiv & \ \prod_{A,B=1}^N\frac{N_{\emptyset \lambda_A}(\bar \mu_B/x_A;\q,\t)N_{\lambda_A \emptyset}(x_A/\mu_B;\q,\t)}{N_{\lambda_A \lambda_B}(x_A/x_B;\q,\t)}~,\label{eq:ZSQCD}
\end{align}
where $\Lambda$ denotes the instanton counting parameter. Let us now partially Higgs the theory by locking the gauge parameters to the (fundamental) flavor ones as follows\footnote{The brane interpretation of such fine tuning is pictorially represented in Fig.\ref{fig:pqDefect}.}
\be
x_A\to x_A^*\equiv \mu_A \t^{r_A^*}\q^{-c_A^*}~,\quad (\ul {r^*},\ul {c^*})\in\mathbb{Z}_{\geq 0}^N\times \mathbb{Z}_{\geq 0}^N~,
\ee
where other equivalent choices are related by permutation of indices. At these points vortex-like configurations open up (we refer to \cite{Aganagic:2013tta} for a discussion in our context), and the Higgsed instanton partition function can be interpreted as the partition function of a bulk-defect coupled system: in our case the 5d bulk turns out to be free (hypers), hence $Z[\text{SQCD}]$ becomes proportional to the vortex-like part of the partition function of the defect. The main novelty w.r.t. simpler Higgsings (e.g. $c^*_A=0$) is that the defect is supported on the intersecting sub-space $[\mathbb{C}\times\mathbb{S}^1]\cup [\mathbb{C}\times\mathbb{S}^1]$ (Fig.\ref{fig:IntersectingDefect}), so that the theory can be described  in the UV (see below) by a pair of 3d $\mathcal{N}=2$ gauge theories on the two orthogonal spaces, with gauge groups $\text{U}(r^*)$ and $\text{U}(c^*)$ respectively and interacting through 1d d.o.f. along the common $\mathbb{S}^1$ at the origin \cite{Gomis:2016ljm,Pan:2016fbl,Nieri:2017ntx,Nieri:2018pev}. 

\begin{figure}[!ht]
\begin{center}
\includegraphics[width=0.4\textwidth]{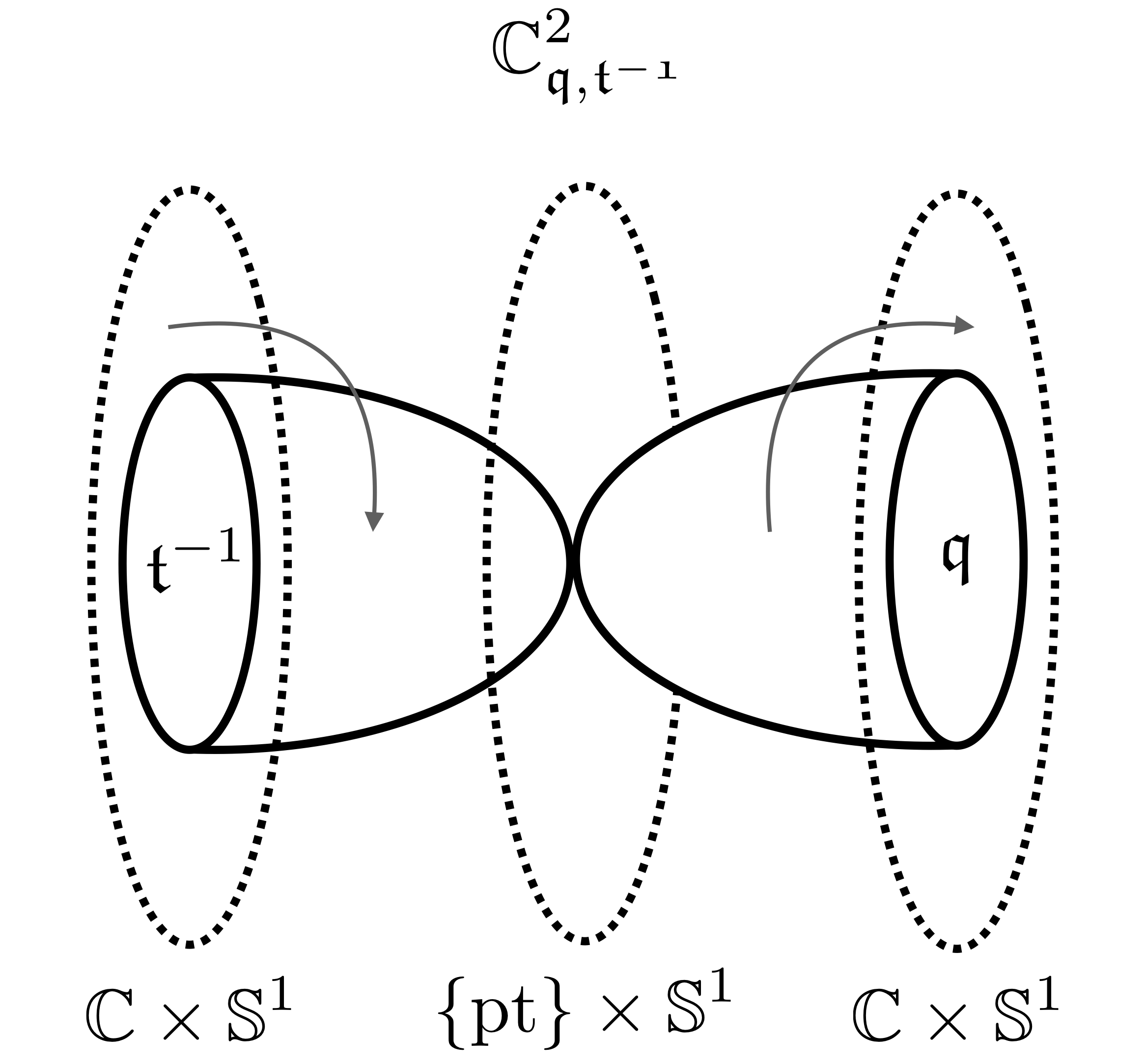}
\end{center}
\caption{Pictorial representation of the support of the defect theory. The cigars intersect along the circle (dashed) that they share at the origin.}
\label{fig:IntersectingDefect}
\end{figure}

\paragraph{The matrix model of the intersecting defect.} In order to see this description emerging from the instanton computation, we recall that the Higgsing condition we are imposing corresponds to the hook truncation discussed around (\ref{eq:NekZeros}). Let us start by focusing on the fixed points labeled by large diagrams only, in which case the previous factorization formulae readily apply to the summands of  (\ref{eq:ZSQCD}) with the trivial identifications $r_A\leftrightarrow r^*_A$, $c_A\leftrightarrow c^*_A$. Also, at these points some recombination between the vector and matter contributions $\Upsilon_\text{x}$, $\Upsilon_\text{f}$ happen, so that we are left with the simpler summands 
\begin{multline}
Z_{\{\lambda_A\}}[\text{SQCD}]\to Z_{\{\lambda_A\}}[\text{SQCD}]^*=\prod_{A,B=1}^N\frac{N_{\emptyset c_A^{r_A}}(\bar \mu_B/x^*_A;\q,\t)N_{c_A^{r_A} \emptyset}(x^*_A/\mu_B;\q,\t)}{N_{c_A^{r_A}c_B^{r_B}}(x^*_A/x^*_B;\q,\t)}\times\\
\times\prod_A (\bar \mu_A/\mu_A)^{|Y^\L |}\,\frac{\Upsilon_\text{v}(\ul z^*_{Y^\R};\q,\t)}{\Upsilon_\text{v}(\ul z^*_{\emptyset^\R};\q,\t)}\frac{\Upsilon_\text{v}(\ul z^*_{Y^\L};\t,\q)}{\Upsilon_\text{v}(\ul z^*_{\emptyset^L};\t,\q)}\times\\
\times\frac{\Upsilon_\text{h}(\eta_\R\ul{\bar\mu},\eta_\R \ul\mu,\ul z^*_{Y^\R};\q)}{\Upsilon_\text{h}(\eta_\R\ul{\bar\mu},\eta_\R \ul\mu,\ul z^*_{\emptyset^\R};\q)}\frac{\Upsilon_\text{h}(1/\eta_\L\ul{\bar\mu},1/\eta_\L \ul\mu,1/\ul z^*_{Y^\L};\t)}{\Upsilon_\text{h}(1/\eta_\L\ul{\bar\mu},1/\eta_\L \ul\mu,1/\ul z^*_{\emptyset^\L};\t)}
 \frac{\Upsilon^\text{1d}_\text{int}(\ul z^*_{Y^\L},\ul z^*_{Y^\R})}{\Upsilon^\text{1d}_\text{int}(\ul z^*_{\emptyset^\L},\ul z^*_{\emptyset^\R})}~,
\end{multline}
where we defined the function
\be
\Upsilon_\text{h}(\ul{\bar u},\ul u,\ul w;\q)\equiv \prod_{i,j}\frac{(\bar u_j/w_i;\q)_\infty}{(u_j/w_i;\q)_\infty}~.
\ee
Note that at the considered locus of parameter space we have
\bes
\begin{align}
\ul z_{Y^\R}\to \ul z^*_{Y^\R}= & \ \{\eta_\R \mu_A \t^{r_A-i}\q^{Y^\R_{Ai}}~, A\in [1,N]~, i \in [1,r_A]\}~,\label{eq:polesR}\\
\ul z_{Y^\L}\to \ul z^*_{Y^\L}= & \ \{\eta_\L \mu_A \q^{i-c_A}\t^{-Y^\L_{Ai}}~,A\in [1,N]~, i \in [1,c_A]\}~.\label{eq:polesL}
\end{align}
\ees
It is then possible to identify the summands above as (some of) the residues of the contour integral/holomorphic block\footnote{We are omitting the constant Cartan contributions from both the vectors and adjoint matter. We will reintroduce them later on.}
\begin{align}\label{eq:qtMatrixModel}
\widehat{Z}[\text{defect}]\equiv & \ \oint\prod_{a=1}^r\frac{\d z_a^\R}{2\pi\i z_a^\R}\, \prod_{b=1}^c\frac{\d z_b^\L}{2\pi\i z_b^\L}\, \check{\Upsilon}^\text{3d}_\q(\ul z^\L;\t)\Upsilon^\text{1d}_\text{int}(\ul z^\L,\ul z^\R){\Upsilon}^\text{3d}_\t(\ul z^\R;\q)~,
\end{align}
where we defined the integrands
\begin{align}
{\Upsilon}^\text{3d}_\t(\ul z^\R;\q)\equiv & \ \Big(\prod_{a=1}^r  z_a^\R\Big)^{\zeta_\R}\Upsilon_\text{v}(\ul z^\R;\q,\t) \Upsilon_\text{h}(\eta_\R\ul{\bar\mu},\eta_\R \ul\mu,\ul z^\R;\q)~,\\
\widehat{\Upsilon}^\text{3d}_\q(\ul z^\L;\t)\equiv & \  \Big(\prod_{b=1}^c  z_b^\L\Big)^{-\zeta_\L}\Upsilon_\text{v}(\ul z^\L;\t,\q) 
\Upsilon_\text{h}(1/\eta_\L\ul{\bar\mu},1/\eta_\L \ul\mu,1/\ul z^\L;\t)~,
\end{align}
provided the following identifications hold
\be
\Lambda\equiv\q^{\zeta_\R}~,\quad\Lambda\prod_A (\bar \mu_A/\mu_A)\equiv\t^{\zeta_\L} \quad \iff \quad  \Lambda\equiv  \q^{\zeta_\R}=\t^{\zeta_\L}\prod_A (\mu_A/\bar\mu_A)~.
\ee
The identification between the holomorphic block integral and the Higgsed instanton partition function is up to normalization by perturbative (1-loop and classical) contributions, which we have not considered yet. The strings of poles (\ref{eq:polesR}) and (\ref{eq:polesL}) attached to the 3d fundamental matter give rise to the residues associated to the large hook diagrams we have focused so far. Remarkably, as detailed e.g. in \cite{Pan:2016fbl}, the contour integral representation of the summands turns out to capture the contribution of the small hook diagrams as well, provided the poles associated to the 1d interaction are considered. We will see how this mechanism explicitly works in an example. The identification of the defect partition function with the Higgsed SQCD partition function is thus explained.

\textit{Remark.} The integrand is multi-valued unless integrality conditions on $\zeta_{\L,\R}$ are imposed. However, as we will recall later on, one should really be inserting suitable $\q$- and $\t$-constants to make the integrand single-valued and keep $\zeta_{\L,\R}$ generic. Then residue calculus can be applied in the form we discussed. This modification does not affect the summands but only the overall normalization. Moreover, the insertion of such $\q$- and $\t$-constants may also be necessary to properly interpret the contour integral via the Jeffrey-Kirwan residue prescription as in \cite{Pan:2016fbl} for the compact case.

As we mentioned, our derivation was based on the chamber $|\q|<1,|\t|<1$, where the $\q$- and $\t$-Pochhammer symbols give rise to the poles and residues we were interested in. However, a similar derivation (in fact, a slightly easier one) holds also in the chamber $|\q<1,|\t^{-1}|<1$, simply by using the reflection properties (\ref{app:neksymm}) differently. While the residues are guaranteed to be the same due to the rational nature of Nekrasov's functions, the matrix model describing the summands is different and given by
\begin{align}\label{eq:qtMatrixModel2}
Z[\text{defect}]\equiv & \ \oint\prod_{a=1}^r\frac{\d z_a^\R}{2\pi\i z_a^\R}\, \prod_{b=1}^c\frac{\d z_b^\L}{2\pi\i z_b^\L}\, {\Upsilon}^\text{3d}_{\q^{-1}}(\ul z^\L;\t^{-1}){\Upsilon}^\text{1d}_\text{int}(\ul z^\L,\ul z^\R){\Upsilon}^\text{3d}_\t(\ul z^\R;\q)~,
\end{align}
where
\begin{align}
{\Upsilon}^\text{3d}_{\q^{-1}}(\ul z^\L;\t^{-1}) \equiv & \ \Big(\prod_{b=1}^c  z_b^\L\Big)^{-\zeta_\L}\Upsilon_\text{v}(\ul z^\L;\t^{-1},\q^{-1}) 
\Upsilon_\text{h}(\eta_\L\ul{\bar\mu},\eta_\L \ul\mu,\ul z^\L;\t^{-1})~,
\end{align}
now with the different identification
\be
\q^{\zeta_\R}\equiv \Lambda\equiv \t^{\zeta_\L}~.
\ee
Note that the selected poles are the same as before. The UV identification of the resulting defect theory can essentially be read from the supersymmetric localized expressions above: the contour integral represents the partition function of a 3d $\mathcal{N}=2$ $\text{U}(r)$ theory on $\mathbb{C}_\q\times\mathbb{S}^1$ coupled to an adjoint chiral of fugacity $\t$, $N$ fundamental/anti-fundamental pairs of chirals with fugacities $\eta_\R\ul\mu$, $\eta_\R\ul{\bar\mu}$ and FI $\zeta_\R$, interacting with another 3d $\mathcal{N}=2$ $\text{U}(c)$ theory on $\mathbb{C}_{\t^{-1}}\times\mathbb{S}^1$ coupled to an adjoint chiral of fugacity $\q$, $N$ fundamental/anti-fundamental pairs of chirals with fugacities $\eta_\L\ul\mu$, $\eta_\L\ul{\bar\mu}$ and FI $-\zeta_\L$. Thus, the defect theory provides  a matter and intersecting generalization \cite{Nieri:2018pev} of the lens space theory \cite{Gukov:2015sna}, as it will be discussed later on. The interaction between the two 3d sub-systems is described by a pair of 1d bi-fundamental chiral multiplets on the common $\mathbb{S}^1$ and superpotential terms, enforcing the relations between the parameters of the two theories. 

As we discuss below, the different expressions across the chambers can be traced back to the appearance of boundary terms (indeed, such subtleties do not arise in the compact backgrounds). A seemingly related analysis has been recently presented in \cite{Lee:2020hfu}. In the following, we will use the notation $\text{U}(r|c)$ to denote the gauge structure of the defect theory: the reason will become manifest momentarily.

\subsection{$(\q,\t)$-deformation of the supermatrix model}\label{subsec:qtMatrixM}

In this section, our main observation is that the models (\ref{eq:qtMatrixModel}) or (\ref{eq:qtMatrixModel2}) provide a $(\q,\t)$-deformation (and holomorphic version) of the  $(r|c)$ supermatrix model, which we are going to argue it arises in the so-called unrefined limit $\t=\q$. Having this specialization in mind, it is easier to discuss the chamber $|\q|<1,|\t|<1$ first, which is compatible with $\t=\q$ within the unit disk. In this case (\ref{eq:qtMatrixModel}) reduces to
\begin{multline}
\widehat{Z}[\text{defect}]\Big|_{\t=\q}= \oint\prod_{a=1}^r\frac{\d z_a^\R}{2\pi\i z_a^\R}\,  \prod_{b=1}^c\frac{\d z_b^\L}{2\pi\i z_b^\L}\,  \Big(\prod_{a}  z_a^\R\Big)^{\zeta-c}\, \Big(\prod_{b}  z_b^\L\Big)^{-\zeta+r}\times\\
\times\frac{\prod_{a\neq a'}(1-z^\R_a/z^\R_{a'})\,\prod_{b\neq b'}(1-z^\L_b/z^\L_{b'})}{(-\q\eta_\L\eta_\R^{-1})^{rc} \prod_{a,b} (1-\q^{-1}\eta_\R\eta_\L^{-1} z^\L_b/z^\R_a)^2}\, \times\\
\times\prod_{A,a}\frac{(\eta_\R \bar\mu_A/z^\R_a;\q)_\infty}{(\eta_\R \mu_A/z^\R_a;\q)_\infty}\prod_{A,b}\frac{(z^\L_b/\eta_\L \bar\mu_A;\q)_\infty}{(z^\L_b/\eta_\L \mu_A;\q)_\infty}\prod_{A,b}\Big(z_b^\L\Big)^{m_A-\bar m_A}~,
\end{multline}
where for convenience we introduced the exponential notation
\be
 \q\equiv\e^{-R\varepsilon}~,\quad \mu_A\equiv\e^{-R\varepsilon m_A}~,\quad \bar \mu_A\equiv\e^{-R\varepsilon \bar m_A}~,
 \ee
 with $\zeta\equiv \zeta_\R=\zeta_\L+\sum_A(m_A-\bar m_A)$. 
Choosing $\q\eta_\L/\eta_\R=1$, the integration measure corresponds indeed to the Cauchy kernel for the eigenvalue integration over $(r|c)$ Hermitian supermatrices
\be
H\equiv \left(\begin{array}{cc}A & B \\ C &D\end{array}\right)\in (r|c)\times (r|c)~,
\ee
where $A\in r\times r$ and $D\in c\times c$ are bosonic, while $B\in r\times c$ and $C\in c\times r$ are fermionic. The integration variables $\ul z^{\R,\L}$ then correspond to the eigenvalues of the block-diagonalized supermatrix 
\be
H\to \left(\begin{array}{cc}H_\R & ~ \\ ~ &H_\L\end{array}\right)~.
\ee

\textit{Remark.} The measure
\be
\Upsilon_\text{v}(\ul z^\R;\q,\t)\Upsilon^\text{1d}_\text{int}(\ul z^\L,\ul z^\R)\Upsilon_\text{v}(\ul z^\L;\t,\q)=\frac{\Delta_{\t}(\ul z^\R;\q)\Delta_{\q}(\ul z^\L;\t)}{\prod_{a,b}(1-\p^{-1/2}z^\L_b/z^\R_a)(1-\p^{-1/2}z^\R_a/z^\L_b)}
\ee
has recently appeared in \cite{atai2021supermacdonald} (see also \cite{sergeev2008deformed}) with reference to the Hilbert space interpretation of super-Macdonald polynomials (defined in a later section), w.r.t. which they are orthogonal and with non-vanishing norms for large diagrams only. Also, instead of taking the unrefined limit, one can set $\t\equiv\q^\beta$ and take the ``conformal limit'' $R\to 0$ while keeping the integration variables fixed. This yields the $\beta$-deformation of the Cauchy kernel
\be
\Upsilon_\text{v}(\ul z^\R;\q,\t)\Upsilon^\text{1d}_\text{int}(\ul z^\L,\ul z^\R)\Upsilon_\text{v}(\ul z^\L;\t,\q)\to 
\frac{\prod_{a\neq a'}(z^\R_a-z^\R_{a'})^\beta\,\prod_{b\neq b'}(z^\L_b-z^\L_{b'})^{1/\beta}}{\prod_{a,b} (z^\L_b- z^\R_a)^2}~,
\ee
which was considered in the context of super-Jack polynomials \cite{Atai_2019}.\footnote{It is worth recalling that we often assume $\eta_\R/\eta_\L=\sqrt{\q\t}$, but more generally $(\eta_\R/\eta_\L)^2=\q\t$ is also a natural choice. In the unitary notation we introduce below, the other branch would convert the $\sinh$ to $\cosh$, which is also a recurrent model in the literature (e.g. the celebrated ABJ(M) matrix model). The measures are related by some simple shifts of the arguments and the potential.} 

The matrix model is supplemented by log potential terms of the type usually encountered in free field approach to 2d CFT (or its $\q$-deformation). Note that the matter-independent part of the potential is compatible with the full supergroup structure since\footnote{The shifts in the exponents appear due to massaging of the measure.}
\bes
\begin{align}
\!\!\!\!\!\Big(\prod_a z^\R_a\Big)^{\zeta+c-r}
\Big(\prod_b z^\L_b\Big)^{-(\zeta+c-r)} \!\!= & \ \text{Sdet}(H)^{\zeta+c-r}~, \quad \text{Sdet}(H)\equiv \frac{\det(H_\R)}{\det(H_\L)}~,\\
\!\!= &\ \e^{-R\varepsilon(\zeta+c-r)\text{Str}(U)}~,\quad \text{Str}(U)\equiv \text{tr}(U_\R)-\text{tr}(U_\L)~,
\end{align}
\ees
where the second line applies to the exponentiated notation $H\equiv \exp(-R\vep U)$ which would give rise to the unitary supermatrix measure.\footnote{We consider the hyperbolic notation, related to the more standard trigonometric one by $R\to \i R$.} However, the matter potential seems to explicitly break the supergroup symmetry down to the bosonic $\text{U}(r)\times \text{U}(c)$. This is indeed true once we fix the contour, that is the pole structure (which we chose so as to reproduce the Higgsed partition function). Suppose instead we change the poles by removing a $\q$-constant, namely we consider the identity 
\begin{multline}
\prod_{A,a}\frac{(\eta_\R \bar\mu_A/z^\R_a;\q)_\infty}{(\eta_\R \mu_A/z^\R_a;\q)_\infty}\prod_{A,b}\frac{(\q\,z^\L_b/\eta_\R \bar\mu_A;\q)_\infty}{(\q\, z^\L_b/\eta_\R \mu_A;\q)_\infty}\prod_{A,b}\Big(z_b^\L\Big)^{m_A-\bar m_A} \,  = \\=\prod_{A,a}\frac{(\eta_\R \bar\mu_A/z^\R_a;\q)_\infty}{(\eta_\R \mu_A/z^\R_a;\q)_\infty}\prod_{A,b}\frac{(\eta_\R \mu_A/z^\L_b;\q)_\infty}{(\eta_\R \bar\mu_A/z^\L_b;\q)_\infty}\times
 \prod_{A,b}\Big(z_b^\L\Big)^{m_A-\bar m_A} \frac{\Theta(\eta_\R \bar\mu_A/z^\L_b;\q)}{\Theta(\eta_\R \mu_A/z^\L_b;\q)}~,
 \end{multline}
and then drop the last factor. This corresponds to changing the initial theory on $\mathbb{C}_\t\times\mathbb{S}^1$ at the boundary torus. The modified potential can be expressed as follows
\be\label{eq:modPot}
\prod_{A,a}\frac{(\eta_\R \bar\mu_A/z^\R_a;\q)_\infty}{(\eta_\R \mu_A/z^\R_a;\q)_\infty}\prod_{A,b}\frac{(\eta_\R \mu_A/z^\L_b;\q)_\infty}{(\eta_\R \bar\mu_A/z^\L_b;\q)_\infty}=\exp\Big(\sum_{n\geq 0}\frac{p_n(\ul t^\R|-\ul t^\L)p_n(\ul z^{-1}_\R|-\ul z^{-1}_\L)}{n}\Big)~,
\ee
where we defined the source and dynamical super-power sums
\be
p_n(\ul u|\ul w)\equiv  p_n(\ul u)-(-1)^np_n(\ul w)~,\quad p_n(\ul u)\equiv \sum_i u_i^n~,
\ee
with
\bes
\begin{align}
\ul t^\R \equiv & \ \{\eta_\R\mu_A\q^{i-1}~,A\in[1,N]~, i\in[1,+\infty)\}~,\\
\ul t^\L\equiv & \  \{\eta_\R\bar \mu\q^{i-1}~,A\in[1,N]~, i\in[1,+\infty)\}~.
\end{align}
\ees
This is sufficient for the potential to be compatible with the full $\text{U}(r|c)$ symmetry. In fact, the expression above allows us to write the supercharacter expansion of the new potential 
\be
\e^{-R\varepsilon(\zeta+c-r)\text{Str}(U)}\, \sum_\lambda \text{Str}_\lambda(T)\, \text{Str}_\lambda(\e^{R\vep U})~,
\ee
where the supermatrix $T$ has eigenvalues $\ul t^{\R,\L}$, and we used that supercharacters are given by the hook or super-Schur polynomials
\be
hs_{\lambda^\vee}(-\ul w|\ul u)=hs_\lambda(\ul u|-\ul w)\equiv\text{Str}_\lambda\Big(\text{diag}(\ul u|\ul w)\Big)~,
\ee
which satisfy the supergroup version of the Cauchy identity
\begin{multline}
\sum_\lambda hs_\lambda(\ul x|-\ul y)hs_\lambda(\ul u|-\ul w)=\frac{\prod_{i,\ell}(1-x_i w_\ell)}{\prod_{i,k}(1-x_i u_k)}\frac{\prod_{j,k}(1-y_j u_k)}{\prod_{j,\ell}(1-y_j w_\ell)}=\\
=\frac{1}{\text{Sdet}\Big(1-\text{diag}(\ul x|\ul y)\otimes\text{diag}(\ul u|\ul w)\Big)}~.
\end{multline}

We can now move to discuss the chamber $|\q|<1, |\t^{-1}|<1$, which is more natural from the gauge theory perspective but also more subtle as far as the unrefined limit is concerned. For convenience, we recall here the partition function  (\ref{eq:qtMatrixModel2}) explicitly
\begin{multline}\label{eq:DefectMMqtinv}
Z[\text{defect}]=\oint\prod_{a=1}^r\frac{\d z_a^\R}{2\pi\i z_a^\R}\, \prod_{b=1}^c\frac{\d z_b^\L}{2\pi\i z_b^\L}\,\Big(\prod_{a}  z_a^\R\Big)^{\zeta_\R}\, \Big(\prod_{b}  z_b^\L\Big)^{-\zeta_\L}\times\\
\times\frac{\Delta_\t(\ul z^\R;\q)\Delta_{\q^{-1}}(\ul z^\L;\t^{-1})}{\prod_{a,b}(1-\q^{-1}\eta_\R\eta_\L^{-1}z^\L_b/z^\R_{a})(1-\t\, \eta_\L\eta_\R^{-1}z^\R_a/z^\L_{b})}\times\\
\times  \prod_{A,a}\frac{(\eta_\R \bar\mu_A/z^\R_a;\q)_\infty}{(\eta_\R \mu_A/z^\R_a;\q)_\infty}\prod_{A,a}\frac{(\eta_\L \bar\mu_A/z^\L_a;\t^{-1})_\infty}{(\eta_\L \mu_A/z^\L_a;\t^{-1})_\infty}~,
\end{multline}
where $\eta_\R/\eta_\L$ can be fixed to $\sqrt{\q\t}$ again. In this case, the limit $\t=\q$ can be taken only at the boundary of the unit disk, and the integrand needs to be defined properly. The measure and interaction term do not pose serious problems because they reduce to the same rational function as before, namely the Cauchy kernel. Moreover, as a formal series, the potential can be directly expanded into supercharacters since its series at $\t=\q$ coincides with the modified potential (\ref{eq:modPot}). This follows from the ``analytic continuation'' from inside to outside the unit circle
\be
(u;\t^{-1})_\infty \to \frac{1}{(\t \, u;\t)_\infty}~,
\ee
which would make sense of the integrand in the unrefined limit. The pole structure changes accordingly, and the previous discussion makes it clear this is again due to modifying the defect theory at the boundary. 

\subsection{Example: $\text{U}(1|1)$ theory}

As a simple yet interesting and illustrative example, let us consider the Abelian $\text{U}(1|1)$ theory, focusing on the 5d SQED ($N=N_\text{f}=N_\text{a.f}=1$) to avoid clutterings. In this case, it is pretty easy to understand the contour and the pole prescription to reproduce the summation over all hook diagrams, large and small. The non-Abelian $\text{U}(r|c)$ Higgsing in the non-Abelian parent 5d theory  follows a similar pattern (for a detailed account  we refer to \cite{Pan:2016fbl}).

\subsubsection{The partition function}

In this example, all the diagrams are obviously large but the trivial one. The former are associated to the poles coming from the purely 3d sector, which are at
\be
z_\R=\eta_\R\mu\q^j~,\quad z_\L=\eta_\L\mu\t^{-i}~,\quad i,j\in\mathbb{Z}_{\geq 0}~.
\ee 
The residues read as
\be
\frac{(\eta_\R\mu)^{\zeta_\R}}{(\eta_\L\mu)^{\zeta_\L}}\frac{(\bar\mu/\mu;\q)_\infty (\bar\mu/\mu;\t^{-1})_\infty}{(\q;\q)_\infty (\t^{-1};\t^{-1})_\infty} \frac{\q^{\zeta_\R\,j}\t^{\zeta_\L\. i}}{(1-\t\,\q^{j}\t^{i})(1-\q^{-1}\q^{-j}\t^{-i})} \frac{(1;\q)_{-j}(1;\t^{-1})_{-i}}{(\bar\mu/\mu;\q)_{-j}(\bar\mu/\mu;\t^{-1})_{-i}}~.
\ee
In order to reproduce the empty diagram (the only small diagram in this example), we can observe that we can take the Left diagram to be empty ($i=0$), but then the Right one needs to ``enter'' inside the $(1,1)$ box, namely it must have a negative entry or add a negative box. This means that the pole must be shifted by one negative $\q$ unit, and this can be accommodated by the interaction term, so that the relevant poles are
\be
z_\L=\eta_\L \mu ~,\quad z_\R=\q^{-1}z_\L\eta_\R\eta_\L^{-1}=\eta_\R\mu\q^{-1}~.
\ee
The residue reads as
\be
Z_\emptyset\equiv \q^{-\zeta_\R}\frac{(\eta_\R\mu)^{\zeta_\R}}{(\eta_\L\mu)^{\zeta_\L}}\frac{(\bar\mu/\mu;\q)_\infty (\bar\mu/\mu;\t^{-1})_\infty}{(\q;\q)_\infty (\t^{-1};\t^{-1})_\infty}\frac{1}{(1-\t\,\q^{-1})(1-\bar\mu/\mu)}~.
\ee
Eventually, using the identification $\q^{\zeta_\R}=\Lambda=\t^{\zeta_\L}$, the partition function evaluates to 
\begin{multline}\label{eq:Zdefect11}
Z[\text{defect}]=Z_\emptyset\Bigg(1+\sum_{i,j\geq 0}(-\Lambda\bar\mu/\mu)^{1+i+j}\frac{(1-\t\, \q^{-1})(1-\mu/\bar\mu)}{(1-\q^{-1}\,\q^{-j}\t^{-i})(1-\t\,\q^{j}\t^{i})}\times\\
\times (-1)^{i+j}\frac{(\q\mu/\bar\mu;\q)_{j}(\t^{-1}\mu/\bar\mu;\t^{-1})_{i}}{(\q;\q)_{j}(\t^{-1};\t^{-1})_{i}}\Bigg)~,
\end{multline}
which coincides with the $\text{U}(1|1)$ Higgsed partition function of the 5d SQED, as it should. The result includes the perturbative contribution, indeed the overall factor can be seen to arise from the classical and 1-loop factors which are not determined by the instanton series (we will briefly comment on such terms in the next section).\footnote{As explained in \cite{Beem:2012mb,Yoshida:2014ssa} (see also \cite{Bullimore:2020jdq} for a recent and exhaustive study), the strange-looking powers involving the FI parameters are due to the non-careful treatment of the classical (mixed-Chern-Simons) terms: one should really be considering the addition of 2d boundary terms which convert them to Theta functions, making the whole integrand genuinely meromorphic with no cuts. A similar comment also holds for the classical 5d bulk action, which requires adding 4d  boundary terms (elliptic Gamma functions) \cite{Nieri:2013vba}. In turn, this specifies the boundary condition and the contour. In the math literature, these contributions are known as elliptic stable envelopes \cite{Aganagic:2016jmx,Aganagic:2017smx,Rimanyi:2019zyi}. Later on, we will need to consider their effect.} 
We can observe that the constant term $\Lambda\, (1-\t\q^{-1})(1-\mu/\bar\mu)$ in the numerator in first line accounts for the rectangular (single box) contribution. By inspection, the summands look like two Macdonald polynomials coupled by some interaction: we will shortly see the reason.

\subsubsection{Intersecting SQED/XYZ duality}

As recalled in a later section, the partition function of Abelian instantons can easily be resummed 
\be
Z[\text{SQED}]=\frac{( \Lambda \bar\mu/x ;\q,\t^{-1})_\infty (\p\, \Lambda x/\mu;\q,\t^{-1})_\infty}{( \Lambda \bar\mu/\mu ;\q,\t^{-1})_\infty (\p\, \Lambda;\q,\t^{-1})_\infty}~.
\ee
Let us now impose the Higgsing condition $x=\mu \t \q^{-1}$. Using
\be
\frac{(u;\q,\t^{-1})_\infty}{(u \t^r\, \q^{-c};\q,\t^{-1})_\infty}=\frac{1}{\prod_{i=1}^r(u\t^{i};\q)_\infty \prod_{j=1}^c(u\q^{-j};\t^{-1})_\infty}\prod_{i=1}^r\prod_{j=1}^c\frac{1}{1-u \t^{i}\q^{-j}}~,
\ee
we have
\be
Z[\text{SQED}]=\frac{(\q\Lambda;\q)_\infty(\t^{-1}\Lambda;\t^{-1})_\infty}{(\q\Lambda \bar\mu/\mu;\q)_\infty(\t^{-1}\Lambda \bar\mu/\mu;\t^{-1})_\infty}\frac{1-\Lambda}{1-\Lambda\bar\mu/\mu}~.
\ee
As explained in \cite{Nieri:2018pev}, the equality between this resummed expression and the series in (\ref{eq:Zdefect11}) can be understood as a consequence of SQED/XYZ mirror duality generalized to intersecting spaces. Mathematically, this identity represents a generalization of the $\q$-binomial theorem
\be
\frac{(v;\q)_\infty}{(u;\q)_\infty}=\sum_{k\geq 0}\frac{(v/u;\q)_k}{(\q;\q)_k}u^k~.
\ee 
As in the ordinary case, we expect this relation to be crucial to extend the proof of the equivalence of partition functions of diverse 3d mirror-like pairs to intersecting spaces. 

\subsubsection{Semiclassical analysis}

This very simple example allows us to study the semiclassical  limit $\t^{-1}\to 1$, namely $\vep_2\to 0$, quite easily. 

\textit{Remark.} Note that this limit explicitly breaks the $\q\leftrightarrow\t^{-1}$ symmetry of the configuration. Therefore, it does not look like a very natural operation in our context, nevertheless it allows us to make contact with similar studies in the literature and appreciate the modifications. Also, for intersecting defects, one should perhaps just talk about NS limit, as the $\vep_2\to 0$ expansion is a semiclassical limit for the theory on the $\mathbb{C}_{\t^{-1}}$ plane, but not for the theory on the $\mathbb{C}_{\q}$ plane, which is still subject to the $\vep_1$-deformation. 

Once the perturbative part is twisted away, we can write the remaining vortex part of the intersecting defect partition function as
\be
Z_\text{v.}[\text{defect}]\equiv \e^{-\frac{1}{R\vep_2}\mathcal{W}(\Lambda;\t^{-1})}\,\Phi(\Lambda;\q)~,
\ee
where
\bes
\begin{align}
\Phi(\Lambda;\q)\equiv & \ \frac{(\Lambda;\q)_\infty}{(\Lambda \bar\mu/\mu;\q)_\infty}~,\\
\mathcal{W}(\Lambda;\t^{-1})\equiv & \  \sum_{k>0}\frac{R\vep_2(1-\bar\mu^k/\mu^k)\t^{-k}\Lambda^k}{k(1-\t^{-k})}=\text{Li}_2(\Lambda)-\text{Li}_2(\Lambda \bar\mu/\mu)+O(\vep_2)~.
\end{align}
\ees
We can thus see that the $\mathbb{C}_{\t^{-1}}$ contribution provides the semiclassical asymptotic, which would be the only contribution in the ordinary non-intersecting setup, whereas the $\mathbb{C}_{\q}$ contribution provides a regular correction in the form of a flat section of a $\q$-connection
\be
\Big(\e^{-R\vep_1\partial_\Lambda}-\frac{1-\Lambda\bar\mu/\mu}{1-\Lambda}\Big)\Phi(\Lambda;\q)=0~.
\ee

\textit{Remark.} The operator annihilating the wavefunction $\Phi$ is equivalent to the the quantum mirror curve of the resolved conifold, as it will be clear from the discussion in the next section. The quantum geometry of refined topological strings was discussed in \cite{Aganagic:2011mi} and, as suggested in \cite{Nekrasov:2009rc}, away from the NS limit the resulting spectral problem becomes non-stationary. Our setup is a further generalization as it not only includes the refinement, but also branes extending along orthogonal supports. 

In the more general non-Abelian case, we can study the NS limit by looking at the saddle point approximation of the block integral (\ref{eq:DefectMMqtinv})
\begin{multline}\label{eq:DefectMMqtinv}
Z[\text{defect}]\simeq \oint\prod_{a=1}^r\frac{\d z_a^\R}{2\pi\i z_a^\R}\,
\frac{\e^{-\frac{1}{R\vep_2}\mathcal{W}(\ul z_\L^*,\Lambda)}\det^{-1/2}\text{Hess}(\mathcal{W}(\ul z_\L^*,\Lambda))}{\prod_{a,b}(1-\q^{-1}\eta_\R\eta_\L^{-1}(z^\L_b)^*/z^\R_{a})(1- \eta_\L\eta_\R^{-1}z^\R_a/(z^\L_{b})^*)}\times\\
\times  \Big(\prod_{a}  z_a^\R\Big)^{-\ln\Lambda/R\vep_1}\, \prod_{A,a}\frac{(\eta_\R \bar\mu_A/z^\R_a;\q)_\infty}{(\eta_\R \mu_A/z^\R_a;\q)_\infty}~,
\end{multline}
where we set
\begin{multline}
\mathcal{W}(\ul z^\L,\Lambda)\equiv \sum_{b\neq b'}\big(\text{Li}_2(z^\L_b/z^\L_{b'})-\text{Li}_2(\q^{-1}z^\L_b/z^\L_{b'})\big)+\\
+\ln\Lambda\sum_{b}\ln z^\L_b+\sum_{A,b}\big(\text{Li}_2(\eta_\L\bar \mu_A/z^\L_b)-\text{Li}_2(\eta_\L\mu_A/z^\L_b)\big)~.
\end{multline}
As usual, the saddle points $\ul z^*_\L$ are determined by the equations $\exp(z^\L_b\partial_{z^\L_b}\mathcal{W}(\ul z^\L,\Lambda))=1$, coinciding with the Bethe equations of a XXZ spin chain according to the gauge/Bethe correspondence  \cite{Nekrasov:2009uh,Nekrasov:2009ui,Nekrasov:2009rc,Nekrasov:2011bc,Nekrasov:2014xaa}
\be
\prod_{b'\neq b}\frac{1-\q z^\L_{b'}/z^\L_b}{1-\q^{-1} z^\L_{b'}/z^\L_b}=\q^{c-1}\Lambda^{-1}\prod_A \frac{1-\eta_\L\mu_A/z^\L_b}{1-\eta_\L\bar\mu_A/z^\L_b}~.
\ee
The main features to observe are that the effective twisted superpotential on the $\mathbb{C}_{\t^{-1}}$ plane gives the asymptotic behaviour of the partition function as expected, while the theory on the $\mathbb{C}_{\q}$ plane is affected by the limit via cancellation of the combined 3d vector-adjoint 1-loop determinants and 1d fugacities determined by the vacuum. As before, one expects the regular 3d-1d contribution to be annihilated by a quantum operator. The systematic gauge theoretic study of such operator identities involve $qq$-characters or Schwinger-Dyson equations \cite{Nekrasov:2015wsu,Haouzi:2020bso}, whose algebraic meaning is recalled in section \ref{sec:4}.  At present, we do not have a clear interpretation from the integrable system viewpoint, however, following \cite{Nekrasov:2017gzb,Nekrasov:2021tik,Jeong:2021bbh}, one may suspect some relation between our system and $\q$KZ equations \cite{Frenkel:1991gx}, perhaps through spectral duality \cite{Fock:1999ae,Mironov:2012ba,Mironov:2013xva}.\footnote{In this regard, it would be interesting to study whether the geometrically defined K-theoretic envelops \cite{Aganagic:2017smx}, playing a crucial role in integral solutions and physically interpretable as 1d d.o.f., can find a gauge theoretic construction via intersecting defects.}

\paragraph{Recap.} Before moving to the next section, let us summarize what we have seen so far. We have implemented the most generic (hook) Higgsing on the instanton partition function of a parent 5d theory. This has allowed us to obtain the partition function of a generalized 3d-1d BPS defect, supported on orthogonal sub-spaces intersecting along a circle at a point. The vortex/instanton summands are rational functions of the 5d $\Omega$-background equivariant parameters $\q,\t^{-1}$, which from the defect viewpoint are either disk or adjoint mass equivariant parameters. We have given two integral representations of the  partition function, akin Coulomb branch localization, one adapted to the chamber $|\q|<1,|\t|<1$, the other one to $|\q|<1,|\t^{-1}|<1$. With a suitable fixed choice of pole structure (boundary d.o.f.), they both reproduce, via residues calculus at the prescribed poles, the partition function of the 5d Higgsed theory.  The integration measure (vector, adjoint and interaction contributions) hints to a deformed supergroup symmetry. Indeed, the resulting matrix-like integrals can be thought of as natural $(\q,\t)$-deformations of the (holomorphic) supermatrix model (Hermitian or unitary depending on the parametrization and allowed potential), which is supposed to arise in the unrefined limit $\t=\q$ only. While the given potential and contour in the chamber $|\q|<1,|\t|<1$ only preserve the bosonic sub-group, the resulting model can be thought of as giving the analytic continuation to the chamber $|\q|<1,|\t^{-1}|<1$, which is manifestly but only formally compatible with the whole super-group symmetry. This situation is analogous, and in fact not completely unrelated, to what happens in the Gaussian supermatrix model: compatibility with the supergroup symmetry requires the Gaussian coefficients to be exactly opposite to one another, but in order to make sense of potentially divergent quantities, one may need to allow a small deformation/analytic continuation away from the supergroup point. 

In the next section, we provide some explanation of the supergroup structure of the defect partition function, which is quite obscure from the purely gauge theoretic viewpoint, from dual M-theory/topological string/Chern-Simons perspectives.

\section{The topological string theory side}\label{sec:3}

In this section, we revisit the previous discussions by embedding our setup in string/M-theory, and we argue by a chain of dualities that the intersecting defect theory should indeed know about  supergroups thanks to the Chern-Simons theory side of open topological strings. This observation explains the appearance of the supergroup structure in the original system, which would seem quite mysterious otherwise (the two 3d $\mathcal{N}=2$ gauge theories making up the intersecting defect are even supported on different space-time components).\footnote{The appearance of supergroups in the context of knot homologies was emphasized in \cite{Gorsky:2013jxa}, where similar brane setups were extensively employed.}

Let us start by describing the 5d $\text{U}(N)$ SQCD on the Coulomb branch in type IIB sting theory through the familiar arrangement of $(p,q)$-branes \cite{Aharony:1997bh,Aharony:1997ju},\footnote{Such configurations actually describe special unitary gauge groups at low energy. However, instanton or topological string computations are usually performed by relaxing the traceless conditions on the gauge generators. The difference is due to the so-called $\text{U}(1)$ factors \cite{Alday:2009aq,Lee:2020hfu}, which are not going to play a significative role in our discussion.} that is a set of 2 NS5's intersecting a stack of $N$ D5's, with the intersections resolved by $(1,1)$-branes (not displayed explicitly in Table \ref{tab:IIBpar}).
\begin{table}[h!]
\begin{center}
\begin{tabular}{c|cccccccccc}
~& $\mathbb{S}^1$ & \multicolumn{2}{c}{$\mathbb{C}_\q$} & \multicolumn{2}{c|}{$\mathbb{C}_{\t^{-1}}$}~&5&6&7&8&9 \\
\hline
2 NS5& --&--&--&--&--&--&~&~&~&~\\
$N$ D5& --&--&--&--&--&~&--&~&~&~\\
\end{tabular}~~~~~.
\end{center}\caption{Our parent setup in IIB language.}\label{tab:IIBpar}
\end{table}

Roughly speaking, the distance between the NS5's encodes the YM coupling constant (instanton counting parameter) while the separations between the D5's encode the Coulomb and (anti-)fundamental mass parameters. The Higgsing procedure considered in the previous section consists to move on the Higgs branch by recombining all the (e.g. right-most) semi-infinite D5's  with the internal ones and resolving the intersections by pulling out the corresponding NS5, while in the process $r$ D3's wrapping the time direction and the $\mathbb{C}_\q$ plane and $c$ D3's wrapping the time direction and the $\mathbb{C}_{\t^{-1}}$ plane in a direction orthogonal to the $(p,q)$-plane are also stretched (Table \ref{tab:IIBdef}).
\begin{table}[h!]
\begin{center}
\begin{tabular}{c|cccccccccc}
~& $\mathbb{S}^1$ & \multicolumn{2}{c}{$\mathbb{C}_\q$} & \multicolumn{2}{c|}{$\mathbb{C}_{\t^{-1}}$}~&5&6&7&8&9 \\
\hline
2 NS5& --&--&--&--&--&--&~&~&~&~\\
$N$ D5& --&--&--&--&--&~&--&~&~&~\\
$r$ D3& --&--&--&~&~&~&~&--&~&~\\
$c$ D3& --&~&~&--&--&~&~&--&~&~
\end{tabular}~~~~~.
\end{center}
\caption{Our defect setup in IIB language.}\label{tab:IIBdef}
\end{table}

The world-volume of the defect theory is supported by the intersecting system of D3's (generalizing the well-known configurations introduced in \cite{Hanany:1996ie,Aharony:1997ju}, see also \cite{Benvenuti:2016wet}), with the FI parameters describing their position relative to the remaining NS5-D5 branes, describing the remaining bulk d.o.f. (free hyper). Since they are stacked, only one independent FI is expected.

\subsection{M-theory perspective}

The IIB description we have just reviewed can be lifted to M-theory, where the $(p,q)$-web we started with is dualized to the degeneration locus of a toric Calabi-Yau 3-fold \cite{Leung:1997tw}. For the sake of simplicity, we focus on the Abelian theory ($N=1$), in which case the dual description of the Higgsed configuration involves M-theory on the well-studied resolved conifold geometry, with the defect corresponding to two stacks of $r$ and $c$ M5's wrapping a toric Lagrangian sub-manifold, as well as the intersecting solid tori $\mathbb{C}_\q\times\mathbb{S}^1$ and  $\mathbb{C}_{\t^{-1}}\times\mathbb{S}^1$ respectively (Table \ref{tab:Mthdef}). The presence of branes wrapping both the $\q$- and $\t$-planes generalizes similar well-studied setups in related contexts (e.g. \cite{Dimofte:2010tz,Witten:2011zz,Gukov:2015sna}). 
\begin{table}[h!]
\begin{center}
\begin{tabular}{c|ccccccccccc}
~&$\mathbb{S}^1$&\multicolumn{2}{c}{$\mathbb{C}_\q$} &\multicolumn{2}{c}{$\mathbb{C}_{\t^{-1}}$}&  \multicolumn{6}{|c}{Resolved conifold} \\
\hline
$r$ M5& --&--&--&~&~&~&~&~&--&--&--\\
$c$ M5& --&~&~&--&--&~&~&~&--&--&--\\
\end{tabular}
\end{center}\caption{Our defect setup in M-theory language.}\label{tab:Mthdef}
\end{table}

Finally, the open part of this geometry can be shown to be dual to the Mikhaylov-Witten brane construction of $\text{U}(r|c)$ Chern-Simons theory \cite{Mikhaylov:2014aoa} (for a recent discussion closer to our context we also refer to \cite{Ferrari:2020avq}), at least in the unrefined limit $\t=\q$. Invoking some generalization of the 3d-3d correspondence \cite{Dimofte:2010tz,Dimofte:2011ju,Dimofte:2011py,Chung:2014qpa,Dimofte:2014ija}, this provides the crucial hint that the intersecting defect should indeed know something about supergroups. Let us observe that the particular (ordinary) case $c=0$ has been extensively studied for a long time: in topological string language, the resolved conifold geometry is simply probed by a stack of Aganagic-Vafa toric branes \cite{Aganagic:2000gs}. In the refined case $\t\neq \q$, one must necessarily consider two types usually dubbed $\q$- and $\t$-branes \cite{Kozcaz:2018ndf}. Note that the distinction is meaningful in the unrefined case as well since one can consider branes supported on orthogonal planes: this generalization is our target in the next subsection.

\subsection{Open/closed duality and  Chern-Simons perspective}

Topological string theory on the resolved conifold geometry is well-known to be dual to ordinary $\text{U}(n)$ Chern-Simons theory on $\mathbb{S}^3$ at large rank \cite{Witten:1992fb}: this is a manifestation of the famous open/closed duality or geometric transition \cite{Gopakumar:1998ki}. In few words, one starts with Chern-Simons theory on $\mathbb{S}^3$, viewed as the zero section of the Calabi-Yau 3-fold $T^*\mathbb{S}^3$ (deformed conifold).  In the A-model description, the theory is supported by a stack of topological branes wrapping the base 3-cycle, hence it can be thought of as an open topological string theory. As the number of branes increases, the size of the 3-cycle shrinks and eventually becomes a singular geometry, which can be resolved by blowing up a $\mathbb{P}^1$ (resolved conifold) leaving behind no branes and hence a closed topological string theory. Wilson loop observables along a knot $K$ in Chern-Simons theory can also be included by following Ooguri-Vafa construction \cite{Ooguri:1999bv}: additional brane probes  wrapping the knot conormal $L_K$ give rise to an open sector, with the probes being essentially unaffected by the geometric transition. In this setup, the topological string partition function computes the generating function (character expansion) of Wilson loops in arbitrary representations in Chern-Simons theory, weighted by source Wilson loops supported on the probes.  These also correspond to the two parameter family of HOMFLY-PT polynomials generalizing the Jones invariant \cite{Witten:1988hf}, where the parameters are related to the size of the $\mathbb{P}^1$ ('t Hooft coupling in Chern-Simons theory) and string coupling  (inverse of renormalized Chern-Simons level). 

Having recalled these basics facts, it is natural to ask what is the Chern-Simons or (refined) open topological string interpretation of the intersecting defect partition function. Following the discussion in the previous subsection, the natural answer is that it should be related to the generating function of loop observables in a $(\q,\t)$-deformation of the supergroup Chern-Simons theory (at large ranks), with the exact supergroup point arising in the unrefined case. 

\textit{Remark.} Note that the closed piece of the geometry is still given by the resolved conifold. The inclusion of more sophisticated probes w.r.t. the usual case allows us to see that a generalization of the open string/Chern-Simons side given by a second gauge factor, eventually combining with the first one into a supergroup, naturally emerges. 

\begin{figure}[!ht]
\begin{center}
\includegraphics[width=0.4\textwidth]{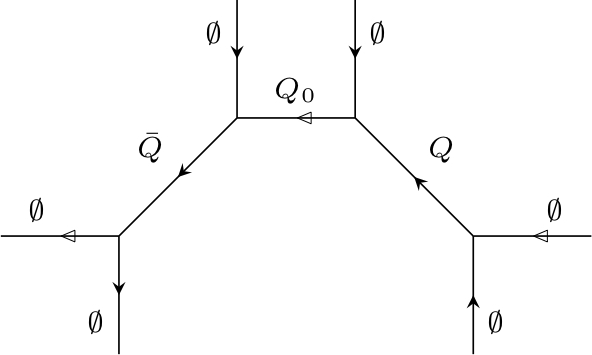}
\end{center}
\caption{The strip geometry engineering the 5d  SQED. The white arrow denotes the preferred (instanton) direction.}
\label{fig:SQEDstrip}
\end{figure}

\paragraph{Standard setup.} In order to support this claim and introduce some notation, let us first review the case of the standard Higgsing, namely $c=0$ (or equivalently $r=0$). Since we are dealing with toric brane probes, we should recover the generating function of the unknot in arbitrary representations (as commented below, this is almost  correct). This follows from the fact that the 5d  SQED instanton partition function (\ref{eq:ZSQCD})
\be\label{eq:ZSQED}
Z[\text{SQED}]\equiv \sum_{\lambda}\Lambda^{|\lambda|}\, Z_{\lambda}[\text{SQED}]\equiv Z_\text{top.}
\ee
coincides with the closed (refined) topological string partition function of the strip geometry (Fig.\ref{fig:SQEDstrip})  \cite{Iqbal:2004ne,Taki:2007dh} upon the identifications 
\be
\Lambda\equiv \p^{-1/2}Q_0\bar Q~,\quad \bar\mu/x\equiv \p^{1/2}\bar Q^{-1}~,\quad x/\mu\equiv \p^{1/2}Q~,
\ee
where the $Q$'s are the exponentiated K{\"a}hler parameters and $\q$ the exponentiated string coupling (in the unrefined limit). When computed through the (refined) topological vertex \cite{Iqbal:2007ii,Awata:2008ed},\footnote{We follow the conventions of the latter reference.} the identification holds up to the perturbative factor
\be
Z_\text{pert.}\equiv \Pi_0(\bar Q \t^\rho,-\q^\rho)\Pi_0( Q \t^\rho,-\q^\rho)~,
\ee 
where $\rho\equiv \{1/2-i, ~i\in[1,+\infty)\}$ and $\Pi_0(\ul x,\ul y)\equiv \prod_{i,j}(1+x_i y_j)$. In the chamber $|\q|<1,|\t^{-1}|<1$, this is simply
\be
Z_\text{pert.}=\frac{1}{(\p^{1/2}\, \bar Q;\q,\t^{-1})_\infty (\p^{1/2}\, Q;\q,\t^{-1})_\infty}~,
\ee
coinciding with the usual gauge theoretic 1-loop determinant of free hypers in the $\Omega$-background $\mathbb{C}^2_{\q,\t^{-1}}\times\mathbb{S}^1$ (possibly up to boundary contributions).

The equivalence with the SQED partition function can be made more explicit by massaging the above expression using the following combinatorial identities 
\bes
\begin{align}\label{eq:NekMac}
N_{\lambda\emptyset}(\p^{1/2}x;\q,\t)=& \ x^{|\lambda|}f_\lambda(\q,\t)\, \frac{P_\lambda(\t^\rho,\p^{-1/2}x^{-1}\t^{-\rho};\q,\t)}{P_\lambda(\t^\rho;\q,\t)}=\frac{P_{\lambda^\vee}(\q^\rho,\p^{1/2}x \q^{-\rho};\t,\q)}{P_{\lambda^\vee}(\q^\rho;\t,\q)}~,\\
N_{\emptyset \lambda}(\p^{1/2}x;\q,\t)=& \ x^{|\lambda|}f_\lambda(\q,\t)^{-1}\, \frac{P_{\lambda^\vee}(\q^\rho,\p^{1/2}x^{-1}\q^{-\rho};\t,\q)}{P_{\lambda^\vee}(\q^\rho;\t,\q)}=\frac{P_{\lambda}(\t^\rho,\p^{-1/2}x \t^{-\rho};\q,\t)}{P_{\lambda}(\t^\rho;\q,\t)}~,\\
N_{\lambda\lambda}(1;\q,\t)=& \ \frac{(\p^{-1/2})^{|\lambda|}}{P_{\lambda^\vee}(-\q^\rho;\t,\q)P_{\lambda}(\t^\rho;\q,\t)}~,
\end{align}
\ees
where $P_\lambda(x;\q,\t)$ is the Macdonald polynomial. We can thus rewrite
\be\label{eq:SQEDMac}
Z_\text{top.}=\sum_\lambda P_{\lambda^\vee}( \q^\rho,\p^{1/2}\bar Q \q^{-\rho};\t,\q)\, (-\p^{-1/2}Q_0)^{|\lambda|}\, P_\lambda(\p^{1/2}Q\t^\rho,  \t^{-\rho};\q,\t)~,
\ee
which is a more traditional form in the (refined) topological string context (related to change of preferred direction) and it can be easily resummed. However, this is not our goal here. Let us now impose the Higgsing condition
\be
Q\to  Q^*=\p^{-1/2}\t^{r}~,\quad r\in\mathbb{Z}_{\geq 0}~.
\ee
Then the last Macdonald polynomial has actually only a finite number of variables and the summation is truncated over diagrams with at most $r$ parts. Therefore 
\be\label{eq:StdGenF1}
Z_\text{top.}\to Z_\text{top.}^*=
 \sum_\lambda P_{\lambda^\vee}(\q^\rho,\p^{1/2}\bar Q\q^{-\rho};\t,\q)\, (-\p^{-1/2}Q_0)^{|\lambda|} \, P_\lambda(\{\t^{i-1/2}\};\q,\t)~,
\ee
where $i\in[1,r]$. We can recognize the factor
\be\label{eq:StdGenF2}
(-1)^{|\lambda|}P_{\lambda^\vee}(\q^{\rho},\p^{1/2}\bar Q\q^{-\rho};\t,\q)\Big|_{\t=\q}=(-1)^{|\lambda|}s_{\lambda^\vee}(\q^{\rho},\bar Q \q^{-\rho})\equiv \bar Q^{|\lambda|/2}\text{dim}_\lambda(\bar Q;\q)
\ee
as the $(\q,\t)$-deformation of the quantum dimension, recovered in the unrefined limit, with spectral parameter $ \bar Q$, that is the (unnormalized) HOMFLY-PT invariant of the unknot. From this perspective, the similar factor
\begin{multline}\label{eq:StdGenF3}
P_\lambda(\{\t^{i-1/2}\};\q,\t)\Big|_{\t=\q} =s_\lambda(\{\q^{i-1/2}\})=s_\lambda(\q^{r+\rho},\q^{-\rho})=\\
(-1)^{|\lambda|}s_{\lambda^\vee}(\q^{\rho},\q^{r-\rho})=\q^{r|\lambda|/2}\text{dim}_\lambda(\q^r;\q)=\q^{\|\lambda^\vee\|^2/2}\prod_{1\leq i<j\leq n}\frac{1-\q^{\lambda_i-\lambda_j+j-i}}{1-\q^{j-i}}
\end{multline}
can be though of as the $(\q,\t)$-deformation of the trace of a background holonomy with Cartan coordinates $\ul x\sim \{\t^{i-1/2}\}$, which in the unrefined limit reduce to the usual character/Schur polynomial. Strictly speaking, the identification of (\ref{eq:StdGenF1}) with the Ooguri-Vafa generating function would be valid if the coordinates were independent, but this is not the case in our setup. From the IIB perspective, this is because in the Higgsing procedure the D3's which support the defect are terminated or compactified by another nearby parallel NS5 and hence dynamical. From the open topological string perspective, it means that those d.o.f. are not really frozen, rather the v.e.v. in the $\text{U}(r)_\kappa$ Chern-Simons theory supported on the probes is taken.\footnote{As usual, the dictionary involves $2\pi\i\ln\q=\kappa+r$.} Therefore, having in mind the matrix model description of the unknot in Chern-Simons theory (e.g. \cite{Brini:2011wi,Morozov:2021zmz} and references therein) or its refined version \cite{Aganagic:2011sg,Aganagic:2012hs}, we have the identification 
\be
s_\lambda(\{\q^{i-1/2}\})\sim \frac{\langle s_\lambda \rangle_{\text{U}(r)_\kappa\text{-CS}}}{\langle 1 \rangle_{\text{U}(r)_\kappa\text{-CS}}}~,
\ee
hence honest sources can be obtained by performing such substitution and omitting the normalized average.\footnote{If one is not willing to perform such operation by hand, a longer strip geometry is needed, i.e. a more complicated (quiver) defect theory is needed.}

\textit{Remark.} In order to compute honest generating functions for toric branes, we could have simply started from the topological vertex applied to the conifold with an open leg and some choice of sources (Macdonald or Schur polynomials). However, since in the generalized setup considered in the following such a choice is not completely obvious, the relation between open and closed Higgsed partition functions we have just worked out proves to be useful.

In order to complete the analysis in this pretty standard setup, let us also recall how the pure Chern-Simons partition function can be computed. Let us consider the limit $Q_0\to 0$, which zooms in the conifold with K{\"a}hler parameter either $\bar Q$ or $Q$. Let us focus on the latter: after Higgsing, the limit decouples the defect from the (free) bulk, and the vortex partition function trivializes (only the empty diagram propagates), implying that the only non-trivial contributions come from the perturbative factors. If we normalize by the $r=0$ configuration (a.k.a. D0-contribution), which in our case is equivalent to extracting the residue at the pole, then we end up with
\be\label{eq:redCS}
\frac{Z_\text{pert.}|_{\p^{1/2}Q=\t^r}}{Z_\text{pert.}|_{\p^{1/2}Q=1}}=\prod_{i=1}^r\frac{1}{(\t^i;\q)_\infty}~,
\ee
which coincides with the refined Chern-Simons partition function on the $\mathbb{S}^3$ \cite{Aganagic:2011sg,Aganagic:2012hs}  before the large $r$ transition in the unreduced normalization \cite{Gukov:2016gkn}. From the refined Chern-Simons perspective, it is natural to include the Cartan contribution $(\t;\q)_\infty^r$, so that the reduced partition function reads as
\be
\prod_{i=1}^r\frac{(\t;\q)_\infty}{(\t^i;\q)_\infty}=\prod_{1\leq i<j\leq r}\frac{(\t^{j-i};\q)_\infty}{(\t^{j-i+1};\q)_\infty}~.
\ee
In the unrefined limit, this simplifies to
\be
\langle 1 \rangle_{\text{U}(r)_\kappa\text{-CS}}\equiv \prod_{1\leq i<j\leq r}(1-\q^{j-i})~,
\ee
which coincides with (the $\q$-analytic part of) the ordinary Chern-Simons partition function on $\mathbb{S}^3$.

\paragraph{Less standard setup.} Now we can essentially repeat the the same steps as above for the general Higgsing condition 
\be
Q\to Q^*=\p^{-1/2}\t^{r}\q^{-c}~,
\ee
with the objective of giving an interpretation to the coefficients of the topological string expansion supporting the supergroup structure. Let us start by recalling the main result for the factorization of Nekrasov's summands when restricted to large hook diagrams with a rectangular plateau $r\times c$, specialized to the Abelian $N=1$ case
\begin{multline}\label{eq:ZSQEDtop}
\Lambda^{|\lambda|}\, Z_{\lambda}[\text{SQED}]= (-\p^{-1/2}Q_0\bar Q)^{rc}\,  Z_\text{rec.}[r,c] \times \\
\times (\p^{-1/2}Q_0\bar Q)^{|Y_\L|+|Y_\R|}\times\frac{(\p\t^{-r} Q \bar Q^{-1})^{|Y_\L|}\, (\q^{-c})^{|Y_\R|}}{Z_\text{int.}[Y_\L,Y_\R]}\times\\
\times\frac{N_{\emptyset Y_\L}(\p^{-1/2}\t^{-r} \bar Q;\t,\q) N_{Y_\L \emptyset}(\p^{-1/2}\t^r Q^{-1};\t,\q)}{N_{Y_\L Y_\L}(1;\t,\q)}\frac{N_{\emptyset Y_\R} (\p^{1/2}\q^{-c}\bar Q^{-1};\q,\t)N_{Y_\R \emptyset}(\p^{1/2}\q^{c} Q;\q,\t)}{N_{Y_\R Y_\R}(1;\q,\t)}~,
\end{multline}
where we set
\bes
\begin{align}
Z_\text{rec.}[r,c]\equiv & \  \q^{r\frac{c^2}{2}}\t^{c\frac{r^2}{2}}\,\p^{rc/2}\prod_{i=1}^r\prod_{j=1}^c \frac{(1-\p^{1/2}Q \t^{1-i}\q^{j-1})(1-\p^{-1/2}\bar Q^{-1}\t^{i-1}\q^{1-j})}{(1-\t^{-1}\, \t^{i-r}\q^{j-c} )(1-\q^{-1}\, \t^{i-r}\q^{j-c})}~,\\
Z_\text{int.}[Y_\L,Y_\R]\equiv &  \  \prod_{i=1}^r\prod_{j=1}^c \frac{(1-\t^{-1}\, \t^{i-r}\q^{j-c}\q^{-Y_{\R i}}\t^{-Y_{\L j}})(1-\q^{-1}\, \t^{i-r}\q^{j-c}\q^{-Y_{\R i}}\t^{-Y_{\L j}})}{(1-\t^{-1}\, \t^{i-r}\q^{j-c})(1-\q^{-1}\, \t^{i-r}\q^{j-c})}~.
\end{align}
\ees
Because of the partial simplifications between these two factors, it is convenient to set
\begin{align}
\widehat{Z}_\text{rec.}[r,c]\equiv& \ \prod_{i=1}^r\prod_{j=1}^c (1-\p^{1/2}Q \t^{1-i}\q^{j-1})(1-\p^{1/2}\bar Q\t^{1-i}\q^{j-1})~,
\end{align}
so that 
\be
\frac{Z_\text{rec.}[r,c]}{Z_\text{int.}[Y_\L,Y_\R]}=\frac{\q^{-r\frac{c^2}{2}}\t^{c\frac{r^2}{2}}\, \q^{c|Y_\R|}\, \t^{r|Y_\L|}\,  \bar Q^{-rc}\, \widehat{Z}_\text{rec.}[r,c]}{\prod_{i=1}^r\prod_{j=1}^c(1-\q^{-1}\, \t^{i-r}\q^{j-c}\q^{-Y_{\R i}}\t^{-Y_{\L j}}) (1-\t\, \t^{r-i}\q^{c-j}\q^{Y_{\R i}}\t^{Y_{\L j}})}
\ee
Let us focus on the Nekrasov factors and let us use again (\ref{eq:NekMac}). Then we may rewrite such factors as
\begin{multline}
(-\p \t^{-r}Q)^{-|Y_\L|}(-\q^{c}\bar Q)^{-|Y_\R|}\times P_{Y^\vee_\L}(\p^{1/2}\t^{-r}\bar Q\t^\rho,\t^{-\rho};\q,\t)P_{Y_\L}(\q^\rho,\p^{1/2}\t^{-r}Q \q^{-\rho};\t,\q)\times\\
\times P_{Y^\vee_\R}(\q^\rho,\p^{1/2}\q^c\bar Q\q^{-\rho};\t,\q)P_{Y_\R}(\p^{1/2}\q^c Q\t^\rho,\t^{-\rho};\q,\t)~.
\end{multline}
Imposing the Higgsing condition, the Macdonald polynomials truncate to a finite number of variables
\bes
\begin{align}
P_{Y_\L}(\q^\rho,\p^{1/2} \t^{-r} Q\q^{-\rho};\t,\q)\to & \ P_{Y_\L}(\{\q^{1/2-j}\};\t,\q)~,\quad i\in[1,r]~,\\
P_{Y_\R}(\p^{1/2}\q^{c} Q\t^\rho,\t^{-\rho};\q,\t)\to & \ P_{Y_\R}(\{\t^{i-1/2}\};\q,\t)~,\quad j\in[1,c]~. 
\end{align}
\ees
Therefore we can eventually write
\begin{multline}\label{eq:MacFac}
\Lambda^{|\lambda|}\, Z_{\lambda}[\text{SQED}]\to  \q^{-r\frac{c^2}{2}}\, \t^{c\frac{r^2}{2}} \, \widehat{Z}_\text{rec.}[r,c]^* \times\\
\times  \t^{r|Y_\L|/2}\, \q^{-c|Y_\R|/2}\, \frac{P_{Y^\vee_\L}(-\p^{1/2}\t^{-r}\bar Q\t^\rho,-  \t^{-\rho};\q,\t)\, P_{Y^\vee_\R}(\q^{\rho},\p^{1/2}\q^c\bar Q\q^{-\rho};\t,\q)}{\prod_{i,j}(1-\q^{-1}\, \t^{i-r}\q^{j-c}\q^{-Y_{\R i}}\t^{-Y_{\L j}})}\times\\
\times (-\p^{-1/2} Q_0)^{rc+|Y_\L|+|Y_\R|}\times \t^{r|Y_\L|/2}\, \q^{-c|Y_\R|/2}\, \frac{P_{Y_\L}(-\{\q^{1/2-j}\};\t,\q)P_{Y_\R}(\{\t^{i-1/2}\};\q,\t)}{\prod_{i,j}(1-\t\, \t^{r-i}\q^{c-j}\q^{Y_{\R i}}\t^{Y_{\L j}})}~.
\end{multline}
In order to interpret these coefficients, we can observe that in the unrefined limit the Macdonald polynomials reduce to Schur polynomials, and as shown in \cite{Moens2003,Eynard:2014rba}\footnote{In the last reference, the computation was made for the $1/\cosh$ interaction (ABJ model), but the same methods apply to our setup, with some caveat about the normalization that we will mention. Later on, we will consider an explicit example.}
\be\label{eq:CShslarge}
\frac{s_{Y_\L}(-\{\q^{1/2-j}\})s_{Y_\R}(\{\q^{i-1/2}\})}{\prod_{i,j}(1- \q^{1+r-i+c-j+Y_{\R i}+Y_{\L j}})}\sim \frac{\langle hs_{\lambda}\rangle_{\text{U}(r|c)_{\kappa}\text{-CS}}}{\langle 1\rangle_{\text{U}(r|0)_{\kappa}\text{-CS}}\langle 1 \rangle_{\text{U}(0|c)_{\kappa}\text{-CS}}}~,
\ee
for a large diagram $\lambda\equiv (c^r)\cup Y_\L^\vee\cup Y_\R$. As we will momentarily see, when combined with part of the contribution from the rectangular part, one can identify the denominator with the supergroup Chern-Simons partition function
\be
\frac{\prod_{i=1}^r\prod_{j=1}^c (1-\p^{1/2}Q^* \t^{1-i}\q^{j-1})}{\langle 1\rangle_{\text{U}(r|0)_{\kappa}\text{-CS}}\langle 1\rangle_{\text{U}(0|c)_{\kappa}\text{-CS}}}\bigg|_{\t=\q}\sim \frac{1}{\langle 1\rangle_{\text{U}(r|c)_{\kappa}\text{-CS}}}~.
\ee
As we explained, in our setup the normalized v.e.v. is automatically taken, and if we wish to consider frozen sources, we may simply ignore it. Actually, some care with this argument is needed because in the unrefined limit the rectangular part is identically zero. Indeed, the naive partition function is divergent and one needs to either work with the unreduced version which can be defined via analytic continuation in the chamber $|\q|<1,|\t|<1$, or consider un-normalized observables. Having said that, the above identification is in line with the supergroup structure and one may be tempted to identify the coefficients of the expansion as a continuous and $(\q,\t)$-deformed version of the quantum superdimension, that is the HOMFLY-PT invariant of the unknot associated to a refined version of supergroup Chern-Simons theory, which we propose in the following. However, note that a large hook diagram with a finite $r\times c$ plateau is actually small in the large rank $\text{U}(m|n)_{\kappa\text{-CS}}$ theory which would arise upon  the $\mathfrak{sl}(n|m)$ specialization $\bar Q\to \bar Q^*=\p^{-1/2} \q^{-m}\t^{n}|_{\t=\q}$, $m,n\in\mathbb{Z}_{\gg0}$. Therefore, even though such coefficients can in principle be compared against an independent matrix model computation, direct analytic checks at finite ranks are expected to be extremely hard to perform. Already in the unrefined limit, exact computations via determinantal representations are possible only for large diagrams: our construction may give a prediction which would be interesting to investigate. In particular, if the mentioned identification is true, then supercharacters (and their deformation) of small representations should somehow factorize after averaging.

\textit{Remark.} Had we started from the unrefined limit from the very beginning, the whole topological expansion would have lost a lot of the structure we are trying to see. The reason is that while it makes sense to consider a number of probes extending along two intersecting subspaces also in that case, the Higgsing condition $Q\to Q^*=\p^{-1/2}\q^{-c}\t^{r}|_{\t=\q}=\q^{r-c}$ would only be sensitive to their difference. This phenomenon is pretty well-known in the context of supermatrix models \cite{AlvarezGaume:1991zc,Yost:1991ht}, and it can be traced back that to the fact that, assuming e.g. $r-c\geq 0$, the Feynman diagrams of the 't Hooft expansion in the $\text{U}(r-c)$ and $\text{U}(r|c)$ models are weighted by the inverse of $r-c=\text{tr}_{\text{U}(r-c)}\mathds{1}=\text{Str}_{\text{U}(r|c)}\mathds{1}$ in both cases. However, this does not mean that the $\text{U}(r|c)$ and the $\text{U}(r-c)$ models are equivalent \cite{Vafa:2014iua}, as is clear after $(\q,\t)$-deformation. Therefore, the refinement is a crucial ingredient in this case (besides regularizing the partition function). Indeed, it is easily seen that in the unrefined limit there are no large diagram contributions as we observed that $\widehat{Z}_\text{rec.}[r,c]$ vanishes in that case: assuming $r-c\geq0$, the resulting truncation coincides with the usual row truncation we have reviewed in the $c=0$ case (with $r\to r-c$); in the opposite regime $r-c<0$, the truncation is on the columns as in the $r=0$ case (with $c\to c-r$). However, the possibility of having ``positive or negative ranks''  may still be taken as a sign of an underlying supergroup theory \cite{Marino:2009jd,Awata:2012jb}.

\subsection{Refined supergroup Chern-Simons theory}

In this section, we study the decoupling limit of the defect from the (free) bulk, which can easily be implemented on the partition function by taking $Q_0\to 0$: then the Higgsed instanton part trivializes, and non-trivial contributions are described by the perturbative part (i.e. the conifold amplitude), which reads as
\be\label{eq:HiggsedConifold}
\frac{Z_\text{pert.}|_{\p^{1/2}Q=\t^r\q^{-c}}}{Z_\text{pert.}|_{\p^{1/2}Q=1}}=\prod_{i=1}^r \frac{1}{(\t^i;\q)_\infty}\prod_{j=1}^c \frac{1}{(\q^{-j};\t^{-1})_\infty}\prod_{i=1}^r\prod_{j=1}^c \frac{1}{(1-\t^i \q^{-j})}~,
\ee
in the chamber $|\q|<1,|\t^{-1}|<1$, or otherwise by analytic continuation. Comparing with (\ref{eq:redCS}), the first two factors can be recognized as two copies of the unreduced refined Chern-Simons partition function on $\mathbb{S}^3$ (the reduced version including the Cartan contributions $(\t;\q)^r_\infty\,(\q^{-1};\t^{-1})^c_\infty$), with refinement parameters $\t$ and  $\q^{-1}$ and (renormalized) coupling constants proportional to $1/\ln\q$ and $-1/\ln\t$ respectively.  The last factor represents some interaction between the two systems and, in view of our discussion, it is natural to assume that their interacting combination will actually be equivalent to some refinement of supergroup Chern-Simons theory. In this subsection, we further support this evidence. 

The first thing to note is that the our analysis naturally suggests the following $(\q,\t)$-deformation of the Chern-Simons matrix model on $\mathbb{S}^3$\footnote{Lens spaces $L(p,1)$ can presumably be described by sending $\q\to \q^{1/p}$ and $\t\to\t^{1/p}$ in the exponent, together with a $\mathbb{Z}_p$ source term.} 
\be\label{eq:refinedSuperCS}
Z_{\q,\t}[\text{CS}]\equiv  \mathcal{N}\oint\prod_{a=1}^r\frac{\d u_a^\R}{2\pi\i}\,  \prod_{b=1}^c\frac{\d u_b^\L}{2\pi\i}\,  \frac{\Delta_\t(\e^{-R\ul u^\R};\q)\, \Delta_{\q^{-1}}(\e^{-R \ul u^\L};\t^{-1})\ \e^{-\frac{R^2}{2\ln\q}\sum_a u_{\R a}^2+\frac{R^2}{2\ln\t}\sum_b u_{\L b}^2}}{\prod_{a,b}\sinh\frac{R}{2}\Big(u^\L_a-u^\R_b-\frac{\vep_+}{2}\Big)\sinh\frac{R}{2}\Big(u^\L_a-u^\R_b+\frac{\vep_+}{2}\Big)}~,
\ee
where we recall the parametrization $\q= \exp(-R\vep_1)$, $\t= \exp(R\vep_2)$, $\p= \exp(-R\vep_+)$, $\vep_+= \vep_1+\vep_2$, while $\mathcal{N}$ is some normalization factor which can be taken to either include (unreduced normalization) or not  (reduced normalization) the Cartan contributions from the adjoint matter in the defect theory
\be
\mathcal{N}\sim \frac{1}{(\t;\q)^r_\infty(\q^{-1};\t^{-1})^c_\infty}~.
\ee
Even if we are using the hyperbolic notation, in order to make contact with (a deformation of) Chern-Simons theory it is convenient to have mind $R\in -\i\mathbb{R}^+$, and consider (small) positive imaginary parts for $\vep_{1,2}$ to define the Macdonald kernels. For $\vep_+=0$, namely at the boundary $\t=\q$ of the unit disk, the integrand of (\ref{eq:refinedSuperCS}) coincides with the integrand of the $\text{U}(r|c)_{\kappa}$ Chern-Simons matrix model with level (up to renormalization) $\kappa\sim 1/\ln\q$, while for $\vep_+\neq 0$ the coupling constants explicitly break the supergroup symmetry. This matrix model is naturally suggested by the intersecting defect picture,\footnote{Had we started with the defect matrix model adapted to the chamber $|\q|<1,|\t|<1$, we would have gotten a similar deformation with $\t^{-1}\to \t$ in the left sector. The resulting model then looks like the deformation of two coupled Gaussian models.} and in fact we can try to derive it from (\ref{eq:DefectMMqtinv}), even though not all the steps are rigorous (we will comment on that in due course). In particular, the derivation naturally suggest an integration contour which is needed to fully define the model. In order to begin with, we slightly massage the interaction term as follows
\begin{multline}
\frac{1}{\prod_{a,b} (1-\p^{-1/2} z^\L_b/z^\R_a)(1-\p^{-1/2} z^\R_a/z^\L_b)}=\\
=\left(-\frac{\p^{1/2}}{4}\right)^{rc}\frac{1}{\prod_{a,b}\sinh\frac{R}{2}\Big(u^\L_a-u^\R_b-\frac{\vep_+}{2}\Big)\sinh\frac{R}{2}\Big(u^\L_a-u^\R_b+\frac{\vep_+}{2}\Big)}~,
\end{multline}
where we used the parametrization $\ul z^{\R,\L}\equiv \exp(-R \ul u^{\R,\L})$. In this way, the interaction term is already reproduced. Then we recall that the classical FI terms are actually to be understood via boundary d.o.f. whose 1-loop determinants are given by Theta functions \cite{Benini:2013nda,Benini:2013xpa,Gadde:2013dda}, namely 
\be
\Big(\prod_a z^\R_a\Big)^{\zeta_\R}\Big( \prod_b z^\L_b\Big)^{-\zeta_\L}\to \prod_{a}\frac{\Theta(\alpha_\R z^\R_a;\q)}{\Theta(\Lambda\alpha_\R z^\R_a;\q)}\prod_{b}\frac{\Theta(\alpha_\L z^\L_b;\t^{-1})}{\Theta(\Lambda\alpha_\L z^\L_b;\t^{-1})}~,
\ee
for some $\alpha_{\R,\L}\in\mathbb{C}^\times$ to be specified momentarily, while we recall the identifications $\q^{\zeta_\R}=\Lambda= \t^{\zeta_\L}$. Apparently, such factors induce full line of poles, but once combined with the matter contributions, $\alpha_{\R,\L}$ can (must) be chosen as to guarantee only two (for each Left/Right sector) separate semi-lines of poles running in opposite directions (the independent contour is still only one). We thus take $\Lambda\alpha_\R=\q/\eta_\R\bar\mu$ and  $\Lambda\alpha_\L=\t^{-1}/\eta_\L\bar\mu$, so that the combined potential reads as
\be
\prod_{a}\frac{\Theta(\alpha_\R z^\R_a;\q)}{(\Lambda\alpha_\R z^\R_a;\q)_\infty (\eta_\R\mu/z_a^\R;\q)_\infty}\prod_{b}\frac{\Theta(\alpha_\L z^\L_b;\t^{-1})}{(\Lambda\alpha_\L z^\L_b;\t^{-1})_\infty (\eta_\L\mu/z_b^\L;\t^{-1})_\infty}~.
\ee
At this point, we are able to take the decoupling limit $\Lambda\sim Q_0\bar Q\to 0$, while keeping $\alpha_{\R,\L}$ fixed. This also requires $\bar\mu\to \infty$, namely $x/\bar\mu\sim \bar Q\to 0$ (indeed, this piece of the geometry is far away). Lastly, we can also consistently send $\mu\to 0$ while keeping  $x/\mu\sim Q\sim \t^{r}\q^{-c}$ fixed. In particular, the latter limit squeezes the semi-lines of poles we were supposed to consider towards the origin, hence we assume that the correct limiting contour is simply a small circle around the origin. We have thus arrived at the contour integral 
\begin{multline}\label{eq:intersectingLens}
Z[\text{defect}]\to \oint\prod_{a=1}^r\frac{\d u_a^\R}{2\pi\i}\,  \prod_{b=1}^c\frac{\d u_b^\L}{2\pi\i}\,  \frac{\Delta_\t(\e^{-R\ul u^\R};\q)\, \Delta_{\q^{-1}}(\e^{-R \ul u^\L};\t^{-1})}{\prod_{a,b}\sinh\frac{R}{2}\Big(u^\L_a-u^\R_b-\frac{\vep_+}{2}\Big)\sinh\frac{R}{2}\Big(u^\L_a-u^\R_b+\frac{\vep_+}{2}\Big)}\times\\
\times \prod_{a}\Theta(\alpha_\R \e^{-R u^\R_a};\q)\prod_{b}\Theta(\alpha_\L \e^{-R u^\L_b};\t^{-1})~,
\end{multline}
where the last line represents the effective contribution from the boundary tori. Note that the definition of $\alpha_{\R,\L}$ requires $\alpha_\R/\alpha_\L=\q\t\eta_\L/\eta_\R$, and with our favourite choice $\eta_\R/\eta_\L=\sqrt{\q\t}$, it is convenient to fix $\alpha_\R\equiv -\q^{1/2}$, $\alpha_\L\equiv -\t^{-1/2}$. In order to complete the justification of our proposal, we now need to use the following identity between integrals of formal series
\be
\sqrt{-\frac{1}{2\pi\i g}}\int_{-\infty}^\infty \d u\, \e^{\i n u-\i u^2 /2 g}=\q^{n^2/2}=\oint_0\frac{\d z}{2\pi\i z}\, z^n\, \vartheta(z;\q)~,\quad \q\equiv \e^{\i g}~,
\ee
for $\text{Im}(g)>0$, where we defined the theta function $\vartheta(z;\q)\equiv \sum_{n\in\mathbb{Z}}\q^{n^2/2}z^n$. When convergence allows the infinite product representation, the relation $(\q;\q)_\infty\Theta(-\q^{1/2}\, z;\q)=\vartheta(z;\q)$ holds true, and then applying this transformation to (\ref{eq:intersectingLens}), we get (\ref{eq:refinedSuperCS}) up to some simple proportionality factor (a similar argument  was also considered in \cite{Okuda:2004mb}).

\textit{Remark.} The partition function (\ref{eq:intersectingLens}) looks like a generalization of the homological block \cite{Gukov:2016gkn} associated to the lens space theory $L(1,1)$ \cite{Gukov:2015sna} (i.e. 3d $\mathcal{N}=2$ Chern-Simons theory with unit level and coupled to an adjoint chiral) via the 3d-3d correspondence, and categorifying the $\text{U}(r)_{\kappa}$ Chern-Simon partition function. This is consistent with our defect theory also being a generalization of the $L(1,1)$ lens space theory to intersecting spaces, and it would be tempting to conclude that its partition function categorifies $\text{U}(r|c)_\kappa$ Chern-Simons invariants (for recent work in this direction, we refer to \cite{Ferrari:2020avq}). One of the subtle points is that in the (non-rigorous) derivation above, we assumed the contour to be around the origin for both Left and Right sectors, while in the Chern-Simons context the unit circle is usually taken. Sometimes this difference is not relevant, for instance when $\beta=\ln\t/\ln\q\in\mathbb{Z}$ in the standard setup $c=0$, as the measure is then polynomial and the origin is the only pole. In our case, some additional care with contour deformations would be needed due to the interaction term and because $\beta,\beta^{-1} \in\mathbb{Z}$ only for $\beta=\pm1$. Let us also observe that the simpler Abelian version of our matrix model has recently appeared in \cite{Liu:2021fsw} in a similar context as ours (intersecting indices and 3d-3d correspondence), without explicit derivation or reference to its supergroup nature though.

\subsection{Example: $\text{U}(1|1)$ theory}

Generally, we do not know how to test our proposal explicitly, as an exact evaluation of this type of matrix integrals seems difficult at present and probably should involve proving some generalization of Macdonald constant term identities first. However, in the simpler case $r=c=1$ (namely the Abelian $\text{U}(1|1)$ theory), things are better and we can try some brute force evaluations. Also, in the unrefined case, such theory can be related to the Alexander polynomial \cite{Rozansky:1992zt} (see also \cite{Eynard:2014rba} are references therein).

\subsubsection{The partition function}

Let us start from the partition function (\ref{eq:intersectingLens}) or (\ref{eq:refinedSuperCS}). In the simpler Abelian theory, we can explicitly show that it is consistent with (\ref{eq:HiggsedConifold}). Indeed, in this case
\be\label{eq:CS11}
Z_{\q,\t}[\text{CS}]\Big|_{r=c=1}\sim \mathcal{N}\oint\frac{\d z^\R}{2\pi\i z^\R}\,  \frac{\d z^\L}{2\pi\i z^\L}\, 
\frac{ \vartheta(z^\R;\q)\, \vartheta(z^\L;\t^{-1})}{(1-\p^{-1/2} z^\L/z^\R)(1-\p^{-1/2} z^\R/z^\L)}~,
\ee
and picking up the constant term of the integrand upon expanding in $z^\R$ and $z^\R/z^\L$, we obtain
\be
Z_{\q,\t}[\text{CS}]\Big|_{r=c=1}\sim \frac{\mathcal{N}}{1-\t\, \q^{-1}}\times\sum_{k\geq 0}\p^{k^2/2}\p^{k/2}(1-\p^{k+1})~.
\ee
The first factor reproduces  (\ref{eq:HiggsedConifold}) in the chosen normalization, while the last series, because of telescopic cancellations, tends to $1$ for $|\p|<1$ (as we assumed in this section).

\subsubsection{Supercharacters}

It is interesting to test more significative observables which are in some sense complete. In the unrefined case, the natural candidates are the supercharacters represented by super-Schur functions in the matrix model. In the Abelian theory, all the representations are large (but the trivial one), so an explicit evaluation is possible. This example is also instrumental for exact higher rank evaluations via determinantal expressions \cite{Eynard:2014rba}. 

The observable we consider is explicitly given by
\be\label{eq:hsss}
hs_\lambda(z_\R|-z_\L)= (-1)^{s'+1}\, z_\R^s\, z_\L^{s'+1}\, (1-z_\R/z_\L)~,
\ee
for $\lambda=(1+s,s')$. Plugging this observable inside the matrix model (\ref{eq:CS11}) and taking the constant term, we obtain
\be
\frac{\langle hs_\lambda(z_\R|-z_\L) \rangle_{\text{U}(1|1)_{\kappa}\text{-CS}}}{\langle 1 \rangle_{\text{U}(1|0)_{\kappa}\text{-CS}}\langle 1 \rangle_{\text{U}(0|1)_{\kappa}\text{-CS}}} \sim \frac{(-1)^{s'}\, \q^{-\frac{s'^2}{2}+\frac{(s+1)^2}{2}}}{1-\q^{1+s+s'}}=\q^{\frac{s(s-1)}{2}-\frac{s'(s'-1)}{2}}\times\frac{(\q^{1/2})^s(-\q^{-1/2})^{s'}}{1-\q^{1+s+s'}}~,
\ee
which agrees with the result (\ref{eq:CShslarge}) up to framing.

It would be interesting to test higher rank observables and/or $(\q,\t)$-deformations. A natural set are for instance  super-Macdonald polynomials \cite{sergeev2008deformed,atai2021supermacdonald}
\be
SP_\lambda(\ul u|\ul v ;\q,\t)\equiv \sum_\mu P_{\lambda/\mu}(\ul u;\q,\t)\, Q_{\mu^\vee}(-\p^{-1/2}\ul v;\t,\q)\xrightarrow{\t=\q}hs_\lambda(\ul u|-\ul v)~,
\ee
where $Q_{\mu^\vee}(\ul v;\t,\q)\equiv \omega_{\q,\t}P_\mu(\ul v;\q,\t)$ is the dual Macdonald polynomial defined by the homomorphism $\omega_{\q,\t}p_n=-p_n(-1)^n(1-\q^n)/(1-\t^{n})$ acting on power sums $p_n$. The relevance of these functions in our context (intersecting 3d $\mathcal{N}=2$ theories and, in the next section, quantum toroidal algebras, was recently pointed out in \cite{Zenkevich:2018fzl,Zenkevich:2019ayk}). For the two variable case, we can write explicit expressions for the fundamental $\lambda=(1)$ representation
\be
SP_{(1)}(z_\R|z_\L;\q,\t)=z_\R-\p^{-1/2}\frac{1-\q}{1-\t}z_\L~,
\ee
and the simplest hook $\lambda=(2,1)$
\begin{multline}
SP_{(2,1)}(z_\R|z_\L ;\q,\t)=\frac{(1-\q)^2}{(1-\t)^2}z_\R (-\p^{-1/2} z_\L)^2\Big(1-\frac{1-\t}{1-\q}\p^{1/2}z_\R/z_\L\Big)+\\
+\frac{\t(1-\p)(1-\q)(1-\q\t)}{(1-\t)^2(1-\t^3)}(-\p^{-1/2}z_\L)^3~.
\end{multline}
In the first line we can recognize the $(\q,\t)$-deformation of (\ref{eq:hsss}), while the second line appears only after deformation. Note that complete factorization seems to be gone for large diagrams too. Generally, we can write the average of a monomial $z_\R^m\, z_\L^n$ as
\be
\frac{\q^{m^2/2}\t^{-n^2/2}\,\q^m\,\t^n}{1-\p^{-1}}\sum_{k\geq 0}\p^{k^2/2}\p^{k/2}(1-\p^{k+1})\, \q^{mk}\, \t^{n k}~.
\ee
Using this formula, we have computed the average of the selected super-Macdonald polynomials and found (to high orders in Mathematica)
\bes
\begin{align}
\langle SP_{(1)}(\q,\t) \rangle_{\q,\t}\sim & \ \mathcal{N}\, \frac{\q^{1/2}}{1-\t}~,\\
\langle SP_{(2,1)}(\q,\t) \rangle_{\q,\t}\sim & \ \mathcal{N}\, \frac{\q^{-1/2}(1-\q)(1-\q\t-\p(1-\t^3))}{(1-\t)^2(1-\t^3)}~.
\end{align}
\ees
Therefore, the factorized result expected from (\ref{eq:MacFac}) and previously observed in the unrefined case is not reproduced (except for the fundamental representation).\footnote{We find non-trivial that the result is simple though.} Unfortunately, with only such polynomials to test, it is difficult to distinguish if this is matter of normalization and accidental cancellations,\footnote{In this regard, let us observe that the quadratic norms computed in \cite{atai2021supermacdonald} contain similar factors we are supposed to reproduce.} or if other combinations are needed. For instance, in the context of ABJ(M) \cite{Drukker:2009hy}, it is known that super-Schur polynomials describe $1/2$ BPS line defects, while Schur polynomials describe the less supersymmetric $1/6$ BPS observables. It is possible that a similar distinction is also necessary in the $(\q,\t)$-deformed setup, giving much more freedom to choose the relevant (perhaps factorized) observables. We do not investigate this direction here.

\section{Quiver $\text{W}_{\q,\t^{-1}}$ algebras and super-instantons}\label{sec:4}

The partition functions of intersecting defects of the type we are considering are known to have a $\text{W}_{\q,\t^{-1}}$ algebra interpretation in the spirit of the BPS/CFT correspondence. In particular, while the defects supported on a single component sub-space (either $\mathbb{C}_\q\times\mathbb{S}^1$ or $\mathbb{C}_{\t^{-1}}\times\mathbb{S}^1$ inside $\mathbb{C}^2_{\q,\t^{-1}}\times\mathbb{S}^1$) can be realized by considering a finite number of screening currents of only one kind \cite{Aganagic:2013tta,Aganagic:2014oia,Aganagic:2015cta}, the defects supported on intersecting subspaces ($[\mathbb{C}_\q\times\mathbb{S}^1]\cup[\mathbb{C}_{\t^{-1}}\times\mathbb{S}^1]$) require both kinds of screening currents of the algebra \cite{Nieri:2017ntx} (for compact spaces, we refer to \cite{Nedelin:2016gwu,Nieri:2017vrb,Nieri:2018pev,Nieri:2018ghd,Cassia:2019sjk,Liu:2021fsw}). Indeed, from the IIB or refined topological string perspective, the two kinds of screening currents are associated to toric $\q$- or $\t$- branes. In the following, we analyze what happens upon sending the number of screening currents of both kinds to infinity: when only one type of screening currents is used, the resulting partition function reproduces the instanton partition function of a parent 5d $\mathcal{N}=1$ theory \cite{Kimura:2015rgi} (see also the review \cite{Kimura:2020jxl}), in agreement with open/closed/open large rank duality; when two types of screening currents are used, one can expect a similar large rank transition. We support this phenomenon by showing that the corresponding state in the $\text{W}_{\q,\t^{-1}}$ algebra side can be identified with a $\beta$-deformation of the time-extended \cite{Losev:2003py,Marshakov:2006ii} instanton partition function of a 5d $\mathcal{N}=1$ supergroup gauge theory \cite{Kimura:2019msw} in the self-dual $\Omega$-background ($\t=\q$). This seems to nicely fit with the discussions in the previous sections, but also poses some puzzles on which we will comment in the final section.

\subsection{$\q$-Virasoro algebra}

For concreteness, we focus here on the $\text{W}_{\q,\t^{-1}}(A_1)$ algebra, namely the $\q$-Virasoro algebra \cite{Shiraishi:1995rp}. Let us consider the Heisenberg algebra generated by oscillators $\{\mb{a}_m, m \in \mathbb{Z}\backslash \{0\} \}$ and zero modes $\mb{P}, \mb{Q}$, with the non-trivial commutation relations 
\begin{align}
 \big[\mb{a}_m, \mb{a}_n\big] &= -\frac{1}{m}(\q^{m/2} - \q^{-m/2})(\t^{-m/2} - \t^{m/2})C^{[m]}(\p)\delta_{m+n, 0}\;, \qquad \big[\mb{P}, \mb{Q} \big] = 2\ ,
\end{align}
where $C^{[m]}(\p)=(\p^{m/2} + \p^{-m/2})$ is the deformed Cartan matrix of the $A_1$ algebra. For any given $\alpha\in\mathbb{C}$, we consider dual Fock modules over the charged Fock vacua $\ket{\alpha}\equiv \e^{\alpha\mb{Q}/2}\ket{0}$ and $\bra{\alpha}\equiv\bra{0}\e^{-\alpha\mb{Q}/2}$ respectively, namely
\be
\mb{P}\ket{\alpha}=\alpha\ket{\alpha} \ , \quad \mb{a}_m\ket{\alpha}=0\ ,\quad \bra{\alpha}\mb{a}_{-m}=0 \ ,\quad m\in\mathbb{Z}_{>0}\ ,
\ee 
with $\braket{0}{0}=1$. 

The $\q$-Virasoro current $\mb{T}(z)\equiv \sum_{m\in\mathbb{Z}}\mb{T}_m z^{-m}$ can be realized as
\be
  \mb{T}(z) \equiv \mb{Y}(\p^{-1/2}z) + \mb{Y}(\p^{1/2}z)^{-1}, \quad \mb{Y}(z) \equiv \ : \exp \Bigg[ \sum_{m \ne 0} \frac{\mb{a}_m  \ z^{-m}}{C^{[m]}(\p)}\Bigg]\q^{\sqrt{\beta}\mb{P}/2}\p^{1/2} :\ , 
  \label{stress-tensor-freeboson}
\ee
where $\beta\in\mathbb{C}$ is such that $\t\equiv \q^\beta$ and the normal ordering $: ~ :$ pushes the positive oscillators and $\mb{P}$ to the right. From the gauge theory perspective, the $\mb{Y}$ operator represents the doubly quantized chiral ring generating function $\mathcal{Y}$ \cite{Nekrasov:2013xda}, while the negative oscillators play the role of higher time variables dual to higher Casimirs (in the 4d cohomological limit). The screening currents of the $\q$-Virasoro algebra have the following free boson representation 
\begin{align}\label{standard-screening-current}
  \mb{S}^{(\pm)}(x)& \equiv  \ :\exp\Bigg[ - \sum_{m \ne 0} \frac{\mb{a}_m \ x^{-m}}{\q_\pm^{m/2} - \q_\pm^{- m/2}} \pm\sqrt{\beta^{\pm 1}}\mb{Q}\pm\sqrt{\beta^{\pm 1}}\mb{P}\ln x\Bigg]: \ ,
\end{align}
where $\mathfrak{q}_+\equiv \q$, $\mathfrak{q}_-\equiv \t^{-1}$. Their defining property is
\be\label{Scurr}
\Big[\mb{T}_m,\mb{S}^{(\pm)}(x)\Big]=\frac{\widehat{T}_{\mathfrak{q}_\pm}-1}{x}\Big( ~\cdots ~\Big) \ ,
\ee
where we have defined a shift operator acting as $\widehat{T}_{\mathfrak{q}_\pm} f(x)=f(\mathfrak{q}_\pm x)$. From the gauge theory viewpoint, this relation is the key for obtaining Ward identities or $qq$-character relations, which can be used to solve for interesting observables \cite{Nekrasov:2015wsu,Nekrasov:2017gzb,Lodin:2018lbz}. The ``OPEs''  of screening currents are as follows
\bes
\begin{align}
\mb{S}^{(\pm)}(x)\mb{S}^{(\pm)}(x')= & \ :\mb{S}^{(+)}(x)\mb{S}^{(+)}(x'):\ \times\nn\\
&\  \times  \exp\Bigg[-\sum_{m>0}\frac{(\q_\mp^{m/2}-\q_\mp^{-m/2})C^{[m]}(\p)(\q_\pm^{1/2} x'/x)^m}{m(1-\q_\pm^m)}\Bigg]\, x^{2\beta^{\pm 1}}~,\\
\mb{S}^{(\mp)}(x)\mb{S}^{(\pm)}(x')=& \  \ :\mb{S}^{(\mp)}(x)\mb{S}^{(\pm)}(x'): \ \frac{(-\p^{1/2}x x')^{-1}}{(1-\p^{-1/2}x/x')(1-\p^{-1/2}x'/x)}\ .
\end{align}
\ees
In the chamber $|\q_\pm|<1$, we can also write
\be
\mb{S}^{(\pm)}(x)\mb{S}^{(\pm)}(x')=\ :\mb{S}^{(\pm)}(x)\mb{S}^{(\pm)}(x'):\ \frac{(x'/x;\q_\pm)_\infty (\p x'/x;\q_\pm)_\infty}{(\q_\pm x'/x;\q_\pm)_\infty(\q_\mp^{-1} x'/x;\q_\pm)_\infty}\, x^{2\beta^{\pm 1}}~,
\ee
while in the chamber $|\q_+|,|\q_-^{-1}|<1$, which we are here mainly interested in, we have
\bes
\begin{align}
\mb{S}^{(+)}(x)\mb{S}^{(+)}(x')= & \ :\mb{S}^{(+)}(x)\mb{S}^{(+)}(x'):\ \frac{(x'/x;\q)_\infty (\p x'/x;\q)_\infty}{(\q x'/x;\q)_\infty(\t x'/x;\q)_\infty}\, x^{2\beta^{\pm 1}}~,\\
\mb{S}^{(-)}(x)\mb{S}^{(-)}(x')= & \ :\mb{S}^{(-)}(x)\mb{S}^{(-)}(x'):\ \frac{(x'/x;\t)_\infty (\p^{-1} x'/x;\t)_\infty}{(\q x'/x;\t)_\infty(\t x'/x;\t)_\infty}\, x^{2\beta^{\pm 1}}~.
\end{align}
\ees

\subsection{Infinitely-many screening charges} 
The screening charges are the operators $\mb{Q}^{(\pm)}$ such that
\be
\Big[\mb{T}_m,\mb{Q}^{(\pm)}\Big]=0 \ .
\ee
Given the property (\ref{Scurr}), a (familiar) definition is  
\be\label{Qint}
\mb{Q}^{(\pm)}\equiv\oint\d x\, c(x;\q_\pm)\ \mb{S}^{(\pm)}(x)\ ,
\ee
for an appropriate choice of integration contour and $\q_\pm$-constant $c(x;\q_\pm)$, namely such that $c(\q_\pm\, x;\q_\pm)=c(x;\q_\pm)$. As an alternative definition, one can consider screening charges defined by Jackson integrals \cite{Kimura:2015rgi}, namely 
\be\label{Qjack}
\mb{Q}^{(\pm)}_{x}\equiv \sum_{k\in\mathbb{Z}}x \mathfrak{q}_\pm^k \ \mb{S}^{(\pm)}(x \mathfrak{q}_\pm^k)\ .
\ee
The latter definition is convenient when considering the insertion of infinitely-many screening charges as there are no explicit integrals to compute, while the additional label (base point $x$) attached to the screening charge is a free modulus. Therefore, we can consider infinitely-many points in the sets (ground configurations)
\bes
\begin{align}
\chi^\R_\emptyset & \ \equiv \{x^\R_A \t^{i-1}\ | A = 1, ..., N \ , i = 1,..., \infty\} \ , \\
\chi^\L_\emptyset & \ \equiv \{x^\L_A \q^{i-1}\ | A = 1, ..., M\ , i = 1,..., \infty\} \ ,
\end{align}
\ees
which we use to define the (un-normalized) operator
\be
\mb{Z}\equiv  \prod^{\succ}_{x\in \chi^\L_\emptyset}\mb{Q}^{(-)}_{x}\prod^{\succ}_{x\in \chi^\R_\emptyset}\mb{Q}^{(+)}_{x}  \ ,
\ee
where $\displaystyle \prod^\succ$ denotes an ordered product.\footnote{\label{ord}We define $\succ$ on $\chi_\emptyset$ by declaring $x_{Ai}\succ x_{Bj}$ if $A>B$, or $i\geq j$ if $A=B$. The ordered product $\prod^\succ$ follows the reverse ordering. Once a suitable chamber is chosen, it can be made to correspond to radial ordering.} This operator involves the summation over the dynamical sets (excited configurations)
\bes
\begin{align}
\chi^\R & \ \equiv \{ (x^\R_A \t^{i-1}\,\q^{-k^\R_{Ai}}\ | A = 1, ..., N \ , i = 1,..., \infty\} \ , \\
\chi^\L & \ \equiv \{x^\L_A\q^{i-1}\, \t^{-k^{\L\vee}_{Ai}}\ | A = 1, ..., M \ , i = 1,..., \infty\} \ ,
\end{align}
\ees
where $k^{\R}$ and $k^{\L\vee}$ can be restricted to $N$-tuples and $M$-tuples of partitions respectively due to the zeros in the OPE when such classification is not satisfied. The ``diagonal'' contributions to the OPE are
\bes
\begin{align}
\prod^{\succ}_{x\in \chi^\R}x\, \mb{S}^{(+)}(x)=& \  \ : \prod^{\succ}_{x\in \chi^\R}\mb{S}^{(+)}(x): \times \prod^{\succ}_{x\in \chi^\R}x^{\sqrt{\beta}(\sqrt{\beta}|\chi^\R|-Q)}\times\nn\\
& \times \prod_{\substack{(x,x')\in \chi^\R\times \chi^\R \\ x\neq x'}} \frac{(x/x';\q)_\infty}{(\t \,x/x';\q)_\infty}\, \prod_{(x\prec x')\in\chi^\R}(x/x')^\beta\frac{\Theta(\t\, x/x';\q)}{\Theta(x/x';\q)}~,\label{eq:OPEplus}\\
\prod^{\succ}_{x\in \chi^\L}x\, \mb{S}^{(-)}(x)=& \  \ : \prod^{\succ}_{x\in \chi^\L}\mb{S}^{(-)}(x): \times \prod^{\succ}_{x\in \chi^\L}x^{\sqrt{\beta^{-1}}(\sqrt{\beta^{-1}}|\chi^\L|+Q)}\times\nn\\
& \times \prod_{\substack{(x,x')\in \chi^\L\times \chi^\L \\ x\neq x'}} \frac{(x/x';\t)_\infty}{(\q\, x/x';\t)_\infty}\, \prod_{(x\prec x')\in\chi^\L}(x/x')^{\beta^{-1}}\frac{\Theta(\q\, x/x';\t)}{\Theta(x/x';\t)}~,\label{eq:OPEminus}
\end{align}
\ees
where $Q\equiv \sqrt{\beta}-1/\sqrt{\beta}$, and we denoted with $|\chi^{\R,\L}|$ the (infinite) cardinality of the set: we will momentarily explain how the infinities arising from the product over infinitely many points can be dealt with. Also, note that the last factors in the OPE, being a $\q$- or $\t$-constant, do not actually depend on the excited configuration. 

In order to make sense of the expressions involving infinite products, we normalize by the ground configuration so that all the factors independent of the dynamical variables cancel out. Part of the surviving OPE factors can be expressed in terms of Nekrasov's functions (i.e. rational functions)
\bes
\begin{align}
&\ \prod_{\substack{(x,x')\in \chi^\R\times \chi^\R \\ x\neq x'}} \frac{(x/x';\q)_\infty}{(\t \,x/x';\q)_\infty}\prod_{\substack{(x,x')\in \chi^\R_\emptyset\times \chi^\R_\emptyset \\ x\neq x'}} \frac{(\t\, x/x';\q)_\infty}{(x/x';\q)_\infty}= \prod_{A,B=1}^N\frac{1}{N_{k_A^\R k_B^\R}(x^\R_{B}/x^\R_A;\q,\t)}~,\label{eq:NekDiagR}\\
&\ \prod_{\substack{(x,x')\in \chi^\L\times \chi^\L \\ x\neq x'}} \frac{(x/x';\t)_\infty}{(\q \,x/x';\q)_\infty}\prod_{\substack{(x,x')\in \chi^\L_\emptyset\times \chi^\L_\emptyset \\ x\neq x'}} \frac{(\q\, x/x';\t)_\infty}{(x/x';\t)_\infty}= \prod_{A,B=1}^M\frac{1}{N_{k_A^{\L\vee} k_B^{\L\vee}}(x^\L_{B}/x^\L_A;\t,\q)}\label{eq:NekDiagL}~,
\end{align}
\ees
and for the last equality we can use the identity
\be
\frac{1}{N_{k_A^{\L\vee} k_B^{\L\vee}}(x^\L_{B}/x^\L_A;\t,\q)}=\frac{(\p^{-1/2}x^\L_A/x^\L_B)^{|k^\L_A|+|k^\L_B|}}{N_{k^\L_A k^\L_B}(x^\L_A/x^\L_B;\q,\t)}\, \frac{f_{k^{\L\vee}_B}(\t,\q)}{f_{k^{\L\vee}_A}(\t,\q)}~.
\ee
It is now clear from the expressions that such normalized OPE factors capture the vector fixed-point contributions of a 5d $\mathcal{N}=1$ $\text{U}(N)\times \text{U}(M)$ gauge theory in the $\Omega$-background $\mathbb{C}^2_{\q,\t^{-1}}\times\mathbb{S}^1$, with equivariant Coulomb parameters $(1/x^\R,x^\L)$. This is not a surprise given the well-known BPS/CFT relations between equivariant K-theory of the moduli space of instantons of quiver gauge theories with unitary groups, and the  free boson representation of quiver $\text{W}_{\q,\t^{-1}}$ algebras. In particular, the $\Omega$-background equivariant parameters $(\varepsilon_1,\varepsilon_2)$ can be identified as follows
\be
\q\equiv \e^{-R\varepsilon_1}~,\quad \t\equiv \e^{R\varepsilon_2}~,\quad \beta\equiv -\varepsilon_2/\varepsilon_1~,
\ee
with $R$ measuring the radius of the fifth compact dimension. However, this is only part of a more interesting story. For later purposes, it is convenient to define the sets
\bes
\begin{align}
\chi^+ & \ \equiv \{ \nu^+_A \t^{i-1}\,\q^{-k^\R_{Ai}}\ | A = 1, ..., N \ , i = 1,..., \infty\}~,\quad \nu^+_A\equiv \eta^+ x^\R_A~, \\
\chi^- & \ \equiv \{\nu^-_A\t^{1-i}\, \q^{k^\L_{Ai}}\ | A = 1, ..., M \ , i = 1,..., \infty\} ~,\quad \nu^-_A\equiv \eta^- x^\L_A~,
\end{align}
\ees
for some $\eta^\pm\in\mathbb{C}^\times$, so that (\ref{eq:NekDiagR}), (\ref{eq:NekDiagL}) become respectively
\bes
\begin{align}
&\ \prod_{\substack{(x,x')\in \chi^+\times \chi^+ \\ x\neq x'}} \frac{(x'/x;\q)_\infty}{(\t \,x'/x;\q)_\infty}\prod_{\substack{(x,x')\in \chi^+_\emptyset\times \chi^+_\emptyset \\ x\neq x'}} \frac{(\t\, x/x';\q)_\infty}{(x/x';\q)_\infty}~,\\
&\ \prod_{\substack{(x,x')\in \chi^-\times \chi^- \\ x\neq x'}} \frac{(x/x';\q)_\infty}{(\t \,x/x';\q)_\infty}\prod_{\substack{(x,x')\in \chi^-_\emptyset\times \chi^-_\emptyset \\ x\neq x'}} \frac{(\t\, x/x';\q)_\infty}{(x/x';\q)_\infty}\times \p^{-N |k^\L|}~.
\end{align}
 \ees

The most interesting and new contributions to the OPE come from the mixed terms
\begin{multline}
:\prod^{\succ}_{x\in \chi^\L}\prod\mb{S}^{(-)}(x): \, :\prod^{\succ}_{x\in \chi^\R}\mb{S}^{(+)}(x): \ = \  \ :\prod^{\succ}_{x\in \chi^\L}\prod\mb{S}^{(-)}(x)\prod^{\succ}_{x\in \chi^\R}\mb{S}^{(+)}(x): \times\\
\times \prod_{(x,x')\in \chi^\L\times\chi^\R}
 \frac{(-\p^{1/2}x x')^{-1}}{(1-\p^{-1/2}x/x')(1-\p^{-1/2}x'/x)}~.
\end{multline}
Once normalized by the ground configuration, we can write the infinite product expressions 
\bes
\begin{align}
\frac{\prod_{(x,x')\in \chi^\L_\emptyset\times\chi^\R_\emptyset}
 (1-\p^{-1/2}x/x')}{\prod_{(x,x')\in \chi^\L\times\chi^\R} (1-\p^{-1/2}x/x')}=& \ \prod_{(x,x')\in \chi^-\times\chi^+}\frac{(\t^{-1} x/x';\q)_\infty}{(x/x';\q)_\infty}\prod_{(x,x')\in \chi^-_\emptyset\times\chi^+_\emptyset}\frac{(x/x';\q)_\infty}{(\t^{-1} x/x';\q)_\infty}~,\\
\frac{\prod_{(x,x')\in \chi^\L_\emptyset\times\chi^\R_\emptyset}
 (1-\p^{-1/2}x'/x)}{\prod_{(x,x')\in \chi^\L\times\chi^\R} (1-\p^{-1/2}x'/x)}=& \ \prod_{(x,x')\in \chi^-\times\chi^+}\frac{(\t\, x'/x;\q)_\infty}{(\t^2\, x'/x;\q)_\infty}\prod_{(x,x')\in \chi^-_\emptyset\times\chi^+_\emptyset}\frac{(\t^2\, x'/x;\q)_\infty}{(\t\, x'/x;\q)_\infty}~,
\end{align}
\ees
provided that $\p^{1/2}\equiv \eta^+/\eta^-$. Now, the crucial observation is that such terms capture a kind of bi-fundamental fixed-point contribution in a 5d $\mathcal{N}=1$ $\text{U}(N)\times\text{U}(M)$ gauge theory, of the precise form that makes the whole OPE factors to match the vector fixed-point contribution of a 5d $\mathcal{N}=1$ $\text{U}(N|M)$ supergroup gauge theory with equivariant parameters $(\nu^+|\nu^-)$, as studied in detail in \cite{Kimura:2019msw}.

In order to complete our analysis, we go back to discussing the possible divergencies due to the infinite number of base points. As we have already mentioned, the normalization by the ground configuration removes most of them. In particular, the non-trivial zero mode contributions read as
\be
\q^{-\sqrt{\beta}|k^\R|(\sqrt{\beta}|\chi^\R|-Q+\mb{P}-|\chi^\L|/\sqrt{\beta})}\t^{-\sqrt{\beta^{-1}}|k^\L|(|\chi^\L|/\sqrt{\beta}+Q-\mb{P}-\sqrt{\beta}|\chi^\R|)}~.
\ee
In order for this to be finite, we renormalize the value of $\mb{P}$ on the charged vacuum as
\be
\mb{P}\to \alpha_\text{ren.}\equiv \xi^\R+Q-\sqrt{\beta}|\chi^\R|+|\chi^\L|/\sqrt{\beta}\equiv-\xi^\L+Q-\sqrt{\beta}|\chi^\R|+|\chi^\L|/\sqrt{\beta}~,
\ee
with $\xi^\R+\xi^\L=0$ and $\xi^{\R,\L}$ finite. Then the effective expansion parameters are finite and given by  $\Lambda$ for the Right sector and $\Lambda^{-1}\p^{-N}$ for the Left sector, with $\Lambda\equiv \q^{-\sqrt{\beta}\xi^\R}=\t^{\xi^\L/\sqrt{\beta}}$. Therefore, the conclusion is that the normalized $\mb{Z}\ket{\alpha_\text{ren.}}$ state computes the (time-extended)  instanton partition function of a $\beta$-deformation of the 5d $\mathcal{N}=1$ $\text{U}(N|M)$ supergroup gauge theory, with the exact supergroup structure emerging only at the unrefined point $\beta=1$ ($\p=1$), where the magnitude squared of the coupling constants is the same and the signs are opposite. This is the type of deformation considered in \cite{Sergeev_2001} in the context of many-body integrable systems associated to supergroups, and further elaborated in \cite{Chen:2020rxu} from the viewpoint of the gauge/Bethe correspondence and Seiberg-Witten integrable systems. Our analysis can be generalized to include matter (e.g. by either shifting the time/oscillator variables or including vertex operators) and arbitrary quiver theories.

\section{Comments and open questions}\label{sec:5}

In this paper, we have argued for a tight relationship or duality between seemingly distant classes of QFT: on the one hand, we find codim-2/4 BPS intersecting defects of QFTs with eight supercharges (specifically, in 5d) based on ordinary gauge groups; on the other hand, there are QFTs supported on ordinary spaces (specifically, 3d) but with supergroup gauge symmetry. The $\Omega$-background on the supersymmetric side induces a deformation on the supergroup side, with the honest supergroup symmetry recovered at the unrefined or self-dual point. In some particular example, we have been able to trace back the supergroup structure of the intersecting defect theory to (a refinement of) supergroup Chern-Simons theory via chains of string dualities. Presumably, this observation can be generalized to more complicated supersymmetric (quiver) gauge theories by invoking an extension of the 3d-3d correspondence or large rank duality. We believe our work is only scratching the surface towards a deeper understanding of generalized gauge theories (either in the sense of the supports or the local symmetries), as there are many open issues and directions worth of future investigations. In the following, we give a brief outline of some of these.

\begin{enumerate}
\item From the purely supersymmetric field theory perspective, it would be desirable to have a clearer ADHM-like understanding of the relations (akin to $\text{d}/\text{d}+2$ dualities \cite{Dorey:2011pa,Chen:2011sj}) between instanton calculus and intersecting defects arising upon Higgsing. Actually, this should also motivate further studies of the defects supported on ordinary single component spaces, as in the case of standard Higgsing (e.g. $c_A=0$ in our notation). This situation is extensively considered in the literature, but the ADHM-like analysis is not so clear except for the simpler (Abelian) configurations (e.g. $r_A\in\{0,1\}$ in our notation, we refer to \cite{Nekrasov:2017rqy,Jeong:2018qpc} for recent discussions), especially in view of the pure vortex analysis in \cite{Fujimori:2012ab,Fujimori:2015zaa}. 

\item While on the supersymmetric side we have worked with 5d-3d-1d coupled systems, most of our discussions can be repeated for the 4d-2d-0d dimensional reduction or the 6d-4d-2d lift, and it would be instructive to work explicitly out similarities and differences. Particularly interesting is the latter case where the dual Calabi-Yau geometry is partially compactified \cite{Hollowood:2003cv,Kimura:2017auj,Bastian:2017ary} and  connections among defects, elliptic algebras \cite{Nieri:2015dts,Iqbal:2015fvd,Kimura:2016dys,Koroteev:2019gqi,Ghoneim:2020sqi} and supergroup deformations may be established along the lines of this paper. Furthermore, given that 6d theories on a torus and 5d circular quiver theories on a circle may be related by IIB S-duality exchanging the rank of the gauge group with the rank of the quiver, it would be interesting to study such phenomenon in the context of supergroups. In the cohomological limit, a relationship between Higgsing, defects and 2d Yang-Mills was explored in \cite{Zhang:2016xqi}, and our analysis should provide a further generalization.

\item As far as the 4d-2d-0d dimensional reduction is concerned \cite{Gomis:2016ljm,Pan:2016fbl,Gomis:2014eya}, intersecting defects were originally considered in the context of the AGT correspondence \cite{Alday:2009aq} (see also \cite{LeFloch:2020uop} for an exhaustive review). In Liouville theory, the simplest type of such defects correspond to the insertion of  arbitrary degenerate operators with momentum $\alpha=-r b/2-c b^{-1}/2$, $r,c\in\mathbb{Z}_{\geq 0}$ (and similarly for Toda), generalizing \cite{Alday:2009fs}. Recently, intersecting defects  have been shown to represent a key ingredient for the field theoretic explanation of many 2d CFT features of supersymmetric gauge theories \cite{Nekrasov:2020qcq,Jeong:2020uxz,Jeong:2021bbh}, including KZ/BPZ equations \cite{Alday:2010vg,Kozcaz:2010yp} and relations to isomonodromic deformation problems \cite{Gamayun:2012ma,Iorgov:2014vla,Bonelli:2016qwg,Teschner:2017rve,Bonelli:2019boe,Bonelli:2019yjd,Bonelli:2021rrg}. In particular, vortex defects  \cite{Bonelli:2011fq} (the type considered in this work) are known (sometimes) to be dual \cite{Frenkel:2015rda,Bullimore:2014awa,Nekrasov:2021tik} to orbifold defects \cite{Kanno:2011fw} (see also \cite{Gorsky:2017hro} for their relations to instantons), a phenomenon algebraically related to quantum Langlands or WZW/Liouville duality \cite{Ribault:2005wp,Hikida:2007tq}. It would be interesting to place our analysis in that context.

\item We showed that correlators involving an infinite number  of $\q$-Virasoro screening currents of both types reproduce a $\beta$-deformation of the partition function of a 5d $\mathcal{N}=1$ supergroup gauge theory. This may seem to be expected in view of our main results and open/closed large rank duality, granted that the intersecting defect theory supported on a finite number of probes is indeed dual to $\text{W}_{\q,\t^{-1}}$ correlators with a finite amount of screenings. However, from the geometric engineering viewpoint, the intersecting defect theory we have focused in this work arises upon a $\text{closed}\to\text{open}$ transition, and the geometry we started with is an ordinary one supporting an ordinary supersymmetric gauge theory. In this sense, the reverse transition $\text{open}\to\text{closed}$ seems to pose some puzzle (at least from the  algebraic engineering perspective), as one naturally gets a (deformation of) supergroup gauge theory rather than an ordinary one. This phenomenon may originate from the mentioned facts that the large rank 't Hooft expansions of ordinary and supergroup gauge theories are only perturbatively equivalent \cite{Vafa:2001qf} (at the supergroup or unrefined point), while the latter may give rise to non-unitary gravitational completions \cite{Vafa:2014iua}. Probably related to these subtle issues, there is also the observation that supergroup gauge theories with eight supercharges (specifically, in 5d) can be realized in string theory by using both positive and negative branes, whereas their partition functions can be computed through networks of topological vertices/anti-vertices \cite{Kimura:2020lmc}. While the fully-fledged dual geometric picture is not completely clear at the moment, it should involve supergroup-type singularities \cite{Dijkgraaf:2016lym} which, in the M-theoretic dual frame, may map to the usual toric Calabi-Yau's but with some extra decoration (accounting for the existence of opposite vertices). It is worth mentioning that Calabi-Yau singularities and supergroups have recently appeared~\cite{Rapcak:2019wzw} in connection to the corner VOAs~\cite{Gaiotto:2017euk}.

\item Talking about non-perturbative effects, we may notice, for instance, that the network of vertices/anti-vertices engineering the 5d $\mathcal{N}=1$ $\text{U}(1|1)$ theory also forms the toric skeleton of the local $\mathbb{P}^1\times\mathbb{P}^1$ geometry describing ordinary Chern-Simons theory on the lens space $L(2,1)$ after geometric transition \cite{Aganagic:2002wv,Halmagyi:2003mm}. By analytic continuation, its partition function is closely related to that of the ABJ(M) model \cite{Marino:2009jd,Awata:2012jb}, which was shown to provide a non-perturbative definition of topological strings on that background \cite{Hatsuda:2013oxa}, crucially relying also on refined information in the NS limit. One may thus wonder about connections between such proposal and our analysis around (refined) supergroups and intersecting defects.

\end{enumerate}

\acknowledgments
The work of TK was supported in part by ``Investissements d'Avenir'' program, Project ISITE-BFC (No.~ANR-15-IDEX-0003), EIPHI Graduate School (No.~ANR-17-EURE-0002), and Bourgogne-Franche-Comt{\'e} region.
The research of FN is supported by the Alexander von Humboldt Foundation and the DESY theory group. FN also thanks the Institut de Math{\'e}matiques de Bourgogne for hospitality during summer 2020, where part of this project was developed under the Procope Mobility Grant 2020 by the French embassy in Germany and the DAAD.

\appendix

\section{Conventions}\label{app:conventions}
In this appendix, we summarize the definitions and some property of the special functions we use throughout the main text.

The (infinite) $\q$-Pochhammer symbol or $\q$-factorial is defined by 
\be
(x;\q)_\infty\equiv\prod_{k\geq 0}(1-\q^k x)~,\quad |\q|<1~.
\ee
Using the representation
\be
(x;\q)_\infty=\e^{-{\rm Li}_2(x;\q)}~,\quad {\rm Li}_2(x;\q)\equiv \sum_{k\geq 1}\frac{x^k}{k(1-\q^k)}~,
\ee
it can be extended to the domain $|\q|>1$ by means of 
\be
(\q x;\q)_\infty\to \frac{1}{(x;\q^{-1})_\infty}~.
\ee
The short Jacobi Theta function is defined by
\be\label{Theta}
\Theta(x;\q)\equiv (x;\q)_\infty(\q x^{-1};\q)_\infty~.
\ee
A useful property is
\be\label{Thetashift}
\frac{\Theta(\q^m x;\q)}{\Theta(x;\q)}=(-x \q^{(m-1)/2})^{-m},\quad 
\frac{\Theta(\q^{-m}x;\q)}{\Theta(x;\q)}=(-x^{-1}\q^{(m+1)/2})^{-m}~.
\ee

The double (infinite) $\q$-factorial is defined by 
\be
(x;\q_1,\q_2)_\infty\equiv \prod_{k\geq 0}(1-\q_1^j \q_2^k x)~,\quad |\q_{1,2}|<1~.
\ee
Using the representation
\be
(x;\q_1,\q_2)_\infty=\e^{-{\rm Li}_3(x;\q_1,\q_2)}~,\quad {\rm Li}_3(x;\q_1,\q_2)\equiv\sum_{k\geq 1}\frac{x^k}{k(1-\q_1^k)(1-\q_2^k)}~,
\ee
it can be extended to other domains by means of
\be\label{pqcont}
(\q_2 x;\q_1,\q_2)_\infty\to \frac{1}{(x;\q_1,\q_2^{-1})_\infty}~.
\ee
A useful property that we used in the main text is
\be
\frac{(x;\q_1,\q_2)_\infty}{(x \q_2^{-r}\, \q_1^{-c};\q_1,\q_2)_\infty}=\frac{1}{\prod_{i=1}^r(x\q_2^{-i};\q_1)_\infty \prod_{j=1}^c(x\q_1^{-j};\q_2)_\infty}\prod_{i=1}^r\prod_{j=1}^c\frac{1}{1-x \q_2^{-i}\q_1^{-j}}~.
\ee

Finally, the super-Schur functions can be defined in terms of the ordinary ones via
\be
hs_\lambda(\ul x|-\ul y)\equiv \sum_\mu s_{\lambda/\mu}(\ul x)\, s_{\mu^\vee}(-\ul y)~,
\ee
where $s_{\lambda/\mu}$ are the skew Schur functions defined as
\be
s_{\lambda/\mu}\equiv \sum_{\nu}c^\lambda_{\mu\nu}s_\nu
\ee
in terms of the Littlewood-Richardson coefficients 
\be
s_\mu s_\nu\equiv\sum_\lambda c^\lambda_{\mu\nu}s_\lambda~.
\ee

\section{Computations}\label{app:computations}
In this appendix, we present a detailed derivation of some technical result used in the main text. 

The starting point is the following rewriting of Nekrasov's function, we can easily be proven using its very definition
\begin{multline}
N_{k_A k_B}(x;\q,\t) =  \ \frac{N_{Y^{\text{L}\vee}_A Y^{\text{L}\vee}_B}(x \t^{r_B-r_A};\q,\t)}{N_{Y^{\text{L}\vee}_A \emptyset}(x \t^{r_B-r_A}\q^{-c_B};\q,\t)N_{\emptyset Y^{\text{L}\vee}_B }(x \t^{r_B-r_A}\q^{c_A};\q,\t)}\times\\
  \times\frac{N_{Y^\text{R}_A Y^\text{R}_B}(x \q^{c_A-c_B};\q,\t)}{N_{Y^\text{R}_A \emptyset}(x \q^{c_A-c_B}\t^{r_B};\q,\t)N_{\emptyset Y^\text{R}_B }(x \q^{c_A-c_B}\t^{-r_A};\q,\t)}\times\\
 \times \underbrace{\prod_{i=1}^{r_A}\prod_{j=1}^{r_B}\frac{(x \t^{j-i};\q)_{c_A-c_B}}{(\t\; x \t^{j-i};\q)_{c_A-c_B}}\prod_{i=1}^{r_A}\prod_{j=1}^{\infty}\frac{(x \t^{r_B} \t^{j-i};\q)_{c_A}}{(\t\; x \t^{r_B} \t^{j-i};\q)_{c_A}}\prod_{j=1}^{r_B}\prod_{i=1}^{\infty}\frac{(x \t^{-r_A} \t^{j-i};\q)_{-c_B}}{(\t\; x \t^{-r_A} \t^{j-i};\q)_{-c_B}}}_{Z_\text{rec.}=N_{c_A^{r_A} c_B^{r_B}}(x;\q,\t)}\times\\
 \times N_{Y^\text{R}_A Y^{\text{L}\vee}_B}(x \t^{r_B}\q^{c_A};\q,\t)N_{Y^{\text{L}\vee}_A Y^\text{R}_B}(x \t^{-r_A}\q^{-c_B};\q,\t)~. 
\end{multline}
Now we apply (\ref{app:neksymm}) to the first line so that
\begin{multline}
\frac{N_{Y^{\text{L}\vee}_A Y^{\text{L}\vee}_B}(x \t^{r_B-r_A};\q,\t)}{N_{Y^{\text{L}\vee}_A \emptyset}(x \t^{r_B-r_A}\q^{-c_B};\q,\t)N_{\emptyset Y^{\text{L}\vee}_B }(x \t^{r_B-r_A}\q^{c_A};\q,\t)}=\\
=\frac{N_{Y^{\text{L}}_A Y^{\text{L}}_B}(x^{-1} \t^{-(r_B-r_A)};\t,\q)}{N_{Y^{\text{L}}_A \emptyset}(x^{-1} \t^{-(r_B-r_A)}\q^{c_B};\t,\q)N_{\emptyset Y^{\text{L}}_B }(x^{-1} \t^{-(r_B-r_A)}\q^{-c_A};\t,\q)}\times \q^{c_A|Y^{\text{L}}_B|-c_B|Y^{\text{L}}_A|}~.
\end{multline}
Note that the last factor disappears when we take the product over all the $A,B$ range. For the last line, we use the following infinite product representation
\begin{align}
N_{\mu^\vee\nu}(x;\q,\t)=& \ \prod_{i,j=1}^\infty\frac{1-\q^{-1}x\, \q^{i-\nu_j}\t^{j-\mu_i}}{1-\q^{-1}x\, \q^{i}\t^{j}}=\nn\\
=& \ \prod_{i=1}^{\ell(\mu)}\prod_{j=1}^{\ell(\nu)}\frac{1-\q^{-1}x\, \q^{i-\nu_j}\t^{j-\mu_i}}{1- \q^{-1}x\, \q^{i}\t^{j}}\,N_{\mu^\vee\emptyset}(x \t^{\ell(\nu)};\q,\t)N_{\emptyset\nu}(x \q^{\ell(\mu)};\q,\t)~,  \\
N_{\mu\nu^\vee}(x;\q,\t)=& \ N_{\mu^\vee\nu}(x^{-1};\t,\q)\times (\p^{-1/2}x)^{|\mu|+|\nu|}f_\nu(\t,\q)/f_{\mu^\vee}(\t,\q)~.
\end{align}
Applying these identities to our case, we have
\begin{align}
N_{Y^{\text{L}\vee}_A Y^\text{R}_B}(x \t^{-r_A}\q^{-c_B};\q,\t)=& \ \prod_{i=1}^{r_B}\prod_{j=1}^{c_A}\frac{1-\q^{-1}x \q^{j-c_B}\t^{i-r_A}\q^{-Y_{Bi}^\text{R}}\t^{-Y_{Aj}^\text{L}}}{1-\q^{-1}x \q^{j-c_B}\t^{i-r_A}}\times\nn\\
& \ \times N_{Y_A^\text{L}\emptyset}(x^{-1}\t^{r_A-r_B}\q^{c_B};\t,\q)N_{\emptyset Y_B^\text{R}}(x \q^{c_A-c_B}\t^{-r_A};\q,\t)\times\nn\\
& \ \times (\p^{-1/2}x \t^{r_B-r_A}\q^{-c_B})^{|Y_A^\text{L}|}/f_{Y^\text{L}}(\t,\q)~,\\
N_{Y^\text{R}_A Y^{\text{L}\vee}_B}(x \t^{r_B}\q^{c_A};\q,\t)=& \ \prod_{i=1}^{r_A}\prod_{j=1}^{c_B}\frac{1-\t^{-1}x^{-1}\t^{i-r_B}\q^{j-c_A}\q^{-Y_{Ai}^\text{R}}\t^{-Y_{Bj}^\text{L}}}{1-\t^{-1}x^{-1}\t^{i-r_B}\q^{j-c_A}}\times\nn\\
& \ \times N_{\emptyset Y_B^\text{L}}(x^{-1} \t^{r_A-r_B}\q^{-c_A};\t,\q) N_{Y_A^\text{R}\emptyset}(x \q^{c_A-c_B}\t^{r_B};\q,\t)\times\nn\\
& \ \times  f_{Y_B^\text{L}}(\t,\q)(\p^{-1/2}x \t^{r_B}\q^{c_A})^{|Y_B^\text{L}|}\q^{c_B|Y_A^\text{R}|}~.
\end{align}
Note that when we take the product over all the $A,B$ range (with $x\to x_A/x_B$), the extra factors which are left over are simply
\be
\p^{-N|\vec Y^\text{L}|}\t^{r|\vec Y^\text{L}|}\q^{c|\vec Y^\text{R}|}~,\quad r=\sum_A r_A~,\quad c=\sum_A c_A~.
\ee
Finally, we can use
\begin{align}
\frac{N_{\mu \nu}(x;\q,\t)}{N_{\mu \emptyset}(x \t^{\ell(\nu)};\q,\t)N_{\emptyset \nu}(x \t^{-\ell(\mu)};\q,\t)}=& \ \prod_{i=1}^{\ell(\mu)}\prod_{j=1}^{\ell(\nu)}\frac{(\t\, x \t^{j-i}\q^{\mu_i-\nu_j};\q)_\infty}{(x \t^{j-i}\q^{\mu_i-\nu_j};\q)_\infty}\frac{( x \t^{j-i};\q)_\infty}{(\t\, x \t^{j-i};\q)_\infty}~,\\
N_{\mu \emptyset}(x;\q,\t)=& \ \prod_{i=1}^{\ell(\mu)}\frac{(\t\, x \t^{-i};\q)_\infty}{(\t\, x \q^{\mu_i}\t^{-i};\q)_\infty}~,\\
N_{\emptyset\nu }(x;\q,\t)=& \ \prod_{i=1}^{\ell(\nu)}\frac{(x \q^{-\nu_i} \t^{i};\q)_\infty}{( x \t^{i};\q)_\infty}~,
\end{align}
to write
\begin{multline}
\prod_{A,B}\frac{1}{N_{k_A k_B}(x_{AB};\q,\t)}\simeq\prod_{A,B}\prod_{i=1}^{r_A}\prod_{j=1}^{r_B}\frac{(x_{AB}\q^{c_A-c_B} \t^{j-i}\q^{Y^\text{R}_{Ai}-Y^\text{R}_{Bj}};\q)_\infty}{(\t\, x_{AB} \q^{c_A-c_B} \t^{j-i}\q^{Y^\text{R}_{Ai}-Y^\text{R}_{Bj}};\q)_\infty}\times\\
\times \prod_{A,B}\prod_{i=1}^{c_A}\prod_{j=1}^{c_B}\frac{(x_{BA} \t^{r_A-r_B} \q^{j-i}\t^{Y^\text{L}_{Ai}-Y^\text{L}_{Bj}};\t)_\infty}{(\q\, x_{BA} \t^{r_A-r_B} \q^{j-i}\t^{Y^\text{L}_{Ai}-Y^\text{L}_{Bj}};\t)_\infty}\times\\
\times \prod_{A,B}\prod_{i=1}^{r_A}\frac{(\t\, x_{AB} \q^{c_A-c_B}\t^{r_B-i}\q^{Y^\text{R}_{Ai}};\q)_\infty}{(x_{BA} \q^{c_B-c_A}\t^{i-r_B} \q^{-Y^\text{R}_{Ai}};\q)_\infty}\prod_{i=1}^{c_A}\frac{(\q x_{BA} \t^{r_A-r_B}\q^{c_B-i}\t^{Y^\text{L}_{Ai}};\t)_\infty}{(x_{AB} \t^{r_B-r_A}\q^{i-c_B}\t^{-Y^\text{L}_{Ai}};\t)_\infty}\times \\
\times \prod_{A,B}\prod_{i=1}^{r_A}\prod_{j=1}^{c_B}\frac{1}{1-\q^{-1}x_{BA} \t^{i-r_B} \q^{j-c_A}\q^{-Y_{Ai}^\text{R}}\t^{-Y_{Bj}^\text{L}}} \frac{1}{1-\t^{-1}x_{BA}\t^{i-r_B}\q^{j-c_A}\q^{-Y_{Ai}^\text{R}}\t^{-Y_{Bj}^\text{L}}}\times\\
\times Z_\text{rec.}^{-1}\times \p^{N |\vec Y^\text{L}|}\t^{-r|\vec Y^\text{L}|}\q^{-c|\vec Y^\text{R}|}~,
\end{multline}
Where $\simeq$ denotes proportionality factor due to empty diagrams (1-loop), and $Z_\text{rec.}$ denotes the contribution from the rectangular part. We now introduce the short-hand notation
\be
z^\text{R}_{Ai}=\eta_{\text{R}}x_A \q^{c_A}\t^{-i}\q^{Y^\text{R}_{Ai}}~,\quad z^\text{L}_{Ai}=\eta_{\text{L}}x_A \t^{-r_A}\q^{i}\t^{-Y^\text{L}_{Ai}}~,
\ee
so that 
\begin{multline}
\prod_{A,B}\frac{1}{N_{k_A k_B}(x_{AB};\q,\t)}\simeq\prod_{A,B}\prod_{i=1}^{r_A}\prod_{j=1}^{r_B}\frac{(z^\text{R}_{Ai}/z^\text{R}_{Bj};\q)_\infty}{(\t\, z^\text{R}_{Ai}/z^\text{R}_{Bj};\q)_\infty}\times\\
\times \prod_{A,B}\prod_{i=1}^{c_A}\prod_{j=1}^{c_B}\frac{(z^\text{L}_{Bj}/z^\text{L}_{Ai};\t)_\infty}{(\q\, z^\text{L}_{Bj}/z^\text{L}_{Ai};\t)_\infty}\times\\
\times \prod_{A,B}\prod_{i=1}^{r_A}\frac{(\t\, z^\text{R}_{Ai}\eta_\text{R}^{-1}x^{-1}_{B} \q^{-c_B}\t^{r_B};\q)_\infty}{(x_{B} \q^{c_B}\t^{-r_B}\eta_\text{R}/z^\text{R}_{Ai} ;\q)_\infty}\prod_{i=1}^{c_A}\frac{(\q x_{B} \t^{-r_B}\q^{c_B}\eta_\text{L}/z^\text{L}_{Ai};\t)_\infty}{(z^\text{L}_{Ai}\eta^{-1}_\text{L}x_{B}^{-1} \t^{r_B}\q^{-c_B};\t)_\infty}\times \\
\times \prod_{A,B}\prod_{i=1}^{r_A}\prod_{j=1}^{c_B}\frac{1}{1-\q^{-1}\eta_\text{R}\eta_\text{L}^{-1}z^\text{L}_{Bj}/z^\text{R}_{Ai}} \frac{1}{1-\t^{-1}\eta_\text{R}\eta_\text{L}^{-1}z^\text{L}_{Bj}/z^\text{R}_{Ai}}\times\\
\times Z_\text{rec.}^{-1}\times \p^{N |\vec Y^\text{L}|}\prod_{A,B}\prod_{i=1}^{r_A}\prod_{j=1}^{c_B} \q^{-j}\t^{-i}\q^{c_A}\t^{r_B}\eta_\text{R}\eta_\text{L}^{-1}z_{Bj}^\text{L}/z_{Ai}^\text{R}~.
\end{multline}

We now include the contribution from (anti-)fundamental matter
\begin{align}
N_{k_A\emptyset}(x/\mu;\q,\t)=& \ \underbrace{\prod_{i=1}^{r_A}(\t^{1-i}\,x/\mu;\q)_{c_A}}_{Z^\text{f}_\text{rec.}}N_{Y^\text{R}_A\emptyset}(\q^{c_A}x/\mu;\q,\t)N_{Y^\text{L}_A\emptyset}(\t^{r_A}\mu/x;\t,\q)\times\nn\\
& \ \times (\p^{1/2}\t^{r_A}\mu/x)^{-|Y^\text{L}_A|}/f_{Y^\text{L}_A}(\t,\q)~,\\
N_{\emptyset k_A}(\bar\mu/x;\q,\t)=& \ \underbrace{\prod_{i=1}^{r_A}\frac{1}{(\t^{i}\,\bar \mu /x;\q)_{-c_A}}}_{Z_\text{rec.}^\text{a.f}} N_{\emptyset Y^\text{R}_A}(\q^{-c_A}\bar\mu/x;\q,\t)N_{\emptyset Y^\text{L}_A}(\t^{-r_A}x/\bar\mu;\t,\q)\times\nn\\
& \ \times (\p^{1/2}\t^{-r_A}x/\bar\mu)^{-|Y^\text{L}_A|}f_{Y^\text{L}_A}(\t,\q)~.
\end{align}
In the case $N_\text{f}=N_\text{a.f}$, after the product over the range of $A,B,f$ (with $x\to x_A$, $\mu\to \mu_f$, $\bar\mu\to\bar\mu_f$), the only left over extra factor is
\be
\p^{-N_\text{f}|\vec Y^\text{L}|}\prod_{f}(\bar\mu_f/\mu_f)^{|\vec Y^\text{L}|}~.
\ee
Therefore, for the case $N_\text{f}=N$, we can finally write
\begin{multline}
\prod_{A,B}\frac{N_{\emptyset k_A}(\bar\mu_B/x_A;\q,\t)N_{\emptyset k_A }(x_A/\mu_B;\q,\t)}{N_{k_A k_B}(x_{AB};\q,\t)}\simeq\prod_{A,B}\prod_{i=1}^{r_A}\prod_{j=1}^{r_B}\frac{(z^\text{R}_{Ai}/z^\text{R}_{Bj};\q)_\infty}{(\t\, z^\text{R}_{Ai}/z^\text{R}_{Bj};\q)_\infty}\times\\
\times \prod_{A,B}\prod_{i=1}^{c_A}\prod_{j=1}^{c_B}\frac{(z^\text{L}_{Bj}/z^\text{L}_{Ai};\t)_\infty}{(\q\, z^\text{L}_{Bj}/z^\text{L}_{Ai};\t)_\infty}\times\\
\times \prod_{A,B}\prod_{i=1}^{r_A}\frac{(\t\, \eta_\text{R}^{-1}\q^{-c_B}\t^{r_B}z^\text{R}_{Ai}/x_B;\q)_\infty}{(\q^{c_B}\t^{-r_B}\eta_\text{R} x_{B} /z^\text{R}_{Ai} ;\q)_\infty}\prod_{i=1}^{c_A}\frac{(\q  \t^{-r_B}\q^{c_B}\eta_\text{L} x_{B}/z^\text{L}_{Ai};\t)_\infty}{(\eta^{-1}_\text{L} \t^{r_B}\q^{-c_B}z^\text{L}_{Ai}/x_B;\t)_\infty}\times \\
\times \prod_{A,B}\prod_{i=1}^{r_A}\frac{(\eta_\text{R}\bar\mu_B /z^\text{R}_{Ai};\q)_\infty}{(\t\,\eta_\text{R}^{-1} z^\text{R}_{Ai}/\mu_B;\q)_\infty}\prod_{i=1}^{c_A}\frac{(\eta_\text{L}^{-1}z^\text{L}_{Ai}/\bar\mu_B;\t)_\infty}{(\q\,\eta_\text{L}\mu_B/ z^\text{L}_{Ai};\t)_\infty}\times\\
\times \prod_{A,B}\prod_{i=1}^{r_A}\prod_{j=1}^{c_B}\frac{1}{1-\q^{-1}\eta_\text{R}\eta_\text{L}^{-1}z^\text{L}_{Bj}/z^\text{R}_{Ai}} \frac{1}{1-\t^{-1}\eta_\text{R}\eta_\text{L}^{-1}z^\text{L}_{Bj}/z^\text{R}_{Ai}}\times\\
\times \frac{Z^\text{f}_\text{rec.}Z^\text{a.f.}_\text{rec.}}{Z_\text{rec.}}\times \prod_{A}(\bar\mu_A/\mu_A)^{|\vec Y^\text{L}|}\times\prod_{A,B}\prod_{i=1}^{r_A}\prod_{j=1}^{c_B} \q^{-j}\t^{-i}\q^{c_A}\t^{r_B}\eta_\text{R}\eta_\text{L}^{-1}z_{Bj}^\text{L}/z_{Ai}^\text{R}~.
\end{multline}
If we now impose the Higgsing condition 
\be
x^*_B/\mu_B=\t^{r^*_B}\q^{-c^*_B}~,
\ee
we have that the summands are zero unless $\ell(Y^\text{R}_A)\leq r^*_A$, $\ell(Y^\text{L}_A)\leq c^*_A$. Therefore, in the above computation we can identify $r^*_A\leftrightarrow r_A$, $c^*_A\leftrightarrow c_A$, and simplify part of the vector contribution with part of the hyper contribution
\begin{multline}
\prod_{A,B}\frac{N_{\emptyset k_A}(\bar\mu_B/x^*_A;\q,\t)N_{\emptyset k_A }(x^*_A/\mu_B;\q,\t)}{N_{k_A k_B}(x^*_{AB};\q,\t)}\simeq\prod_{A,B}\prod_{i=1}^{r_A}\prod_{j=1}^{r_B}\frac{(z^\text{R}_{Ai}/z^\text{R}_{Bj};\q)_\infty}{(\t\, z^\text{R}_{Ai}/z^\text{R}_{Bj};\q)_\infty}\times\\
\times \prod_{A,B}\prod_{i=1}^{c_A}\prod_{j=1}^{c_B}\frac{(z^\text{L}_{Bj}/z^\text{L}_{Ai};\t)_\infty}{(\q\, z^\text{L}_{Bj}/z^\text{L}_{Ai};\t)_\infty}\times\\
\times \prod_{A,B}\prod_{i=1}^{r_A}\frac{(\eta_\text{R}\bar\mu_B /z^\text{R}_{Ai};\q)_\infty}{(\eta_\text{R} \mu_{B} /z^\text{R}_{Ai} ;\q)_\infty}\prod_{i=1}^{c_A}\frac{(\eta_\text{L}^{-1}z^\text{L}_{Ai}/\bar\mu_B;\t)_\infty}{(\eta^{-1}_\text{L} z^\text{L}_{Ai}/\mu_{B};\t)_\infty}\times \\
\times \prod_{A,B}\prod_{i=1}^{r_A}\prod_{j=1}^{c_B}\frac{1}{1-\q^{-1}\eta_\text{R}\eta_\text{L}^{-1}z^\text{L}_{Bj}/z^\text{R}_{Ai}} \frac{1}{1-\t^{-1}\eta_\text{R}\eta_\text{L}^{-1}z^\text{L}_{Bj}/z^\text{R}_{Ai}}\times\\
\times \frac{Z^\text{f}_\text{rec.}Z^\text{a.f.}_\text{rec.}}{Z_\text{rec.}}\times \prod_{A}(\bar\mu_A/\mu_A)^{|\vec Y^\text{L}|}\times\prod_{A,B}\prod_{i=1}^{r_A}\prod_{j=1}^{c_B} \q^{-j}\t^{-i}\q^{c_A}\t^{r_B}\eta_\text{R}\eta_\text{L}^{-1}z_{Bj}^\text{L}/z_{Ai}^\text{R}~.
\end{multline}

\bibliographystyle{utphys}
\providecommand{\href}[2]{#2}\begingroup\raggedright\endgroup

\end{document}